\documentclass[12pt]{article}
\usepackage{epsfig}
\usepackage{graphicx}
\usepackage{multicol}
\renewcommand{\theequation}{\thesection.\arabic{equation}}

\def\npb#1#2#3{    {\it Nucl. Phys. }{\bf B #1} (#2) #3}
\def\plb#1#2#3{    {\it Phys. Lett. }{\bf B #1} (#2) #3}

\def\prd#1#2#3{    {\it Phys. Rev. }{\bf D #1} (#2) #3}

\def\prep#1#2#3{   {\it Phys. Rep. }{\bf #1} (#2) #3}

\def\ibid#1#2#3{   {\it ibid. }{\bf #1} (#2) #3}
\def\jhep#1#2#3{   {\it JHEP  }{\bf #1} (#2) #3}

\newcommand{\gsim}{\stackrel{>}{_\sim}}
\newcommand{\ba}{\begin{array}}
\newcommand{\ea}{\end{array}}
\newcommand{\be}{\begin{equation}}
\newcommand{\ee}{\end{equation}}
\newcommand{\bea}{\begin{eqnarray}}
\newcommand{\eea}{\end{eqnarray}}
\newcommand{\beq}{\begin{equation}}
\newcommand{\eeq}{\end{equation}}

\renewcommand\Re{\mbox{Re}}
\renewcommand\a{\alpha}

\newcommand{\cO}{{\cal O}}

\newcommand{\no}{\nonumber}
\newcommand{\Cnew}{C^{\rm new}}
\newcommand{\mbMS}{\overline{m}_b}
\newcommand{\mcpole}{m_c}
\newcommand{\mbpole}{m_b}
\newcommand{\mqpole}{m_{b,c}}

\begin{document}

\thispagestyle{empty}
\begin{flushright}
CERN--TH/2003-132\\
SLAC-PUB-10265\\
MCTP-03-54\\

\end{flushright}

\vspace*{1.5cm}
\centerline{\Large\bf The rare decay  $B \rightarrow X_s \ell^+\ell^-$ to NNLL 
precision}
\vspace*{0.2cm}
\centerline{\Large\bf for arbitrary dilepton invariant mass  }

\vspace*{2cm}
\centerline{ {\large\bf  A.~Ghinculov,$^{a}$\footnote{~Presently with Merrill Lynch 
- Corporate Risk Management.}
 T.~Hurth,$^{b,c,}$\footnote{~Heisenberg Fellow.}
 G.~Isidori,$^{d}$ Y.-P.~Yao$^{e}$} }

\bigskip

\begin{center}
{\it ${}^a$~Department of Physics, University of Rochester\\
         Rochester, NY 14627, USA }

\vspace{0.3cm}

{\it ${}^b$~Theoretical Physics Division, CERN, CH-1211 Geneva 23, Switzerland}\\

\vspace{0.3cm}

{\it ${}^c$~SLAC, Stanford University, Stanford, CA 94309, USA}\\

\vspace{0.3cm}

{\it ${}^d$~INFN, Laboratori Nazionali di Frascati, I-00044 Frascati, Italy}\\

\vspace{0.3cm}

{\it ${}^e$~Michigan Center for Theoretical Physics\\  University of 
Michigan, Ann Arbor  MI 48109-1120, USA}
\end{center}

\vspace*{1.5cm}

\centerline{\large\bf Abstract}
\begin{quote}
We present a new phenomenological analysis of the inclusive 
rare decay $B \rightarrow X_s \ell^+\ell^-$. In particular, 
we present the first calculation of the NNLL contributions 
due to the leading two-loop matrix elements, evaluated 
for arbitrary dilepton invariant mass.
This allows to obtain the first NNLL estimates of the dilepton mass spectrum 
and the lepton forward-backward asymmetry in the high $ M^2_{\ell^+ \ell^-}$
region, and to provide an independent check of previously published 
results in the low $ M^2_{\ell^+ \ell^-}$  region.
The numerical impact of these NNLL corrections in the high-mass region
($ M^2_{\ell^+ \ell^-} > 14.4~{\rm GeV}^2$) amounts to $-13\%$
in the integrated rate, and leads to a reduction of the scale 
uncertainty to $\pm 3\%$. The impact of non-perturbative contributions 
in this region is also discussed in detail. 

\end{quote}
\vfill
\newpage
\pagenumbering{arabic}

\setcounter{equation}{0}
\section{Introduction}
\label{intro}
It has been more than a quarter of a century since the basic blocks were laid
to build up the Standard Model (SM).  Since then a wealth of high precision 
calculations and measurements has been performed and compared.  
The SM passed all the tests.  
And yet, it is universally accepted that there must be 
something more than what has been so brilliantly conceived and verified.  
Our undaunted quests to understand fundamental problems 
such as the flavour families, the dark matter or  
quantum gravity, have led to several proposals of new models.
Short of direct evidences about the new degrees of freedom of the theory, 
it remains a fact that some of the most constructive 
and definitive work to search for extensions of the SM
lies in the cold hard world of higher and higher precision tests:
the systematic search for discrepancies between theoretical 
estimates and experimental data in quantities particularly 
sensitive to the symmetry structure of the model.

\medskip

Various possibilities of precision tests have been considered, 
some of which derive from flavour-changing neutral-current processes (FCNC).  
The key argument here is that neutral flavour-violating 
processes can occur only via loops, within the SM, and 
leads to tight constraints on possible new sources of flavour-symmetry 
breaking terms, even if these appear well above the electroweak
scale (see e.g.~Refs.~\cite{MFV,Borzumati,Review}). 
The new $B$ factories, BELLE and BABAR,
are  providing us now with more and more high statistics data, which will 
carry FCNC tests  to the next precision level. 
For our part, 
we shall now focus on an inclusive process of 
the type  $B\to X_s \ell^+ \ell^-$, where in principle 
$X_s$ stands for anything with a single strange quark.  
Since the emphasis is on precision, we must work 
in a region where uncertainties can be minimized or at least reliably bounded.  
As long as the (sub)energy scales are much larger 
than the characteristic $\Lambda_{\rm QCD}$ we can treat 
the problem on hand primarily at a parton level, where perturbative calculations are 
justifiably deployed.
Non-perturbative effects either must be under control or even can 
be added with confidence.  
These requirements 
call for our avoiding the regions where the dilepton 
invariant mass is near the $c\bar c$
resonances.

\medskip

After such considerations, it is natural to divide the $B \to X_s \ell^+ \ell^-$ 
data into two distinct
sets, depending on the squared invariant mass ($ M^2_{\ell^+ \ell^-} \equiv q^2$)
of the dilepton pair:
$$1\ \mbox{GeV}^2< q^2 < 6 \, \mbox{GeV}^2 \ \ \mbox{[low]}; 
\qquad\qquad  q^2 > 14.4 \ \mbox{GeV}^2  \ \ \mbox{[high]}.$$
These two windows are in fact complementary to each other, both experimentally and 
theoretically, on the one hand because of identification and detection 
efficiency and event rates, on 
the other because of non-perturbative and perturbative effects, as well as 
parametric  uncertainties.  We shall give results 
that will cover these two regions.

\medskip

It was recognized some time ago that, because of the mixing structure
and because of the very heavy top mass, rates for processes such as $B \to X_s \gamma$
and $B \to X_s \ell^+ \ell^-$ could be sizable. It was also realized that these 
processes are ideal 
candidates for a short-distance analysis based on the operator product expansion. 
Partonic QCD corrections to the pure electroweak amplitudes 
can be investigated systematically, via renormalization group improvement, 
and are found to change the rates significantly. With the help of the heavy-quark 
expansion one can also reliably assess non-perturbative contributions, 
which are found to be small for sufficiently inclusive observables. 
As one may infer from above, there are essentially three ingredients which go into 
a high-accuracy calculation of these FCNC processes.
The first one is the perturbative evaluation of the partonic amplitudes at the electroweak scale,
which can be translated into the initial conditions for an effective theory 
description based on a local Hamiltonian,
${\cal H}_{{\rm eff}} = \sum C_i \, {\cal O}_i$, of dimension-6 operators. 
The second ingredient deals with the subsequent evolution of ${\cal H}_{{\rm eff}}$
down to the actual scale  of the process, namely the running of the effective 
coupling constants $C_i$ -- which tentamounts 
to dressing them with QCD -- from $M_{W, t}$ to $m_b$
via renormalization group improvement. The last step concerns the evaluation 
of a combination of process-dependent matrix elements corresponding to  
the specific observable:  a combination of the type
$\int d \Gamma | \langle X_s \ell^+ \ell^- | \sum C_i  {\cal O}_i | B \rangle |^2$ in our case. 
This last step includes hard QCD corrections, which
again can be computed perturbatively, and the attendant bremsstrahlung and soft gluon effects, 
since QCD is a massless gauge theory. It also 
includes the non-perturbative hadronization of $b$ and $s$ quarks, 
which can be analyzed with the help of the heavy-quark expansion.
Each of the three steps must be taken to matching 
orders of accuracy in powers of the  strong coupling constants $\alpha_s$, with 
renormalization group resummations whenever necessary.
Within  the present paper, our central focus will be on the last step (hadronic matrix elements) 
and we will give expressions for the  differential rate in the dilepton mass 
and the integrated rates over 
the two windows. We already published results for the forward-backward asymmetry \cite{Adrian1,Results}, 
where we  detailed our calculation of the bremsstrahlung contributions. 

\medskip

One may view $B \to X_s \ell^+ \ell^-$ as an extension of $B \to X_s \gamma$, in which the virtual 
photon can take on a whole spectrum of mass. This picture is somewhat simplistic and 
rather misleading. Indeed, to describe the former process we need extra electroweak operators, 
which are sensitive to different aspects of the mechanisms of electroweak-
and flavor-symmetry breaking. This is a very important feature, 
which could allow to detect a non-standard effect in  $B \to X_s \ell^+ \ell^-$
even in absence of deviations from the SM in $B \to X_s \gamma$.
On the other hand, there are certainly several common points in the theoretical 
analysis of these two processes. In particular, similarly to the  $B \to X_s \gamma$ case,  
also in $B \to X_s \ell^+ \ell^-$ sizable corrections to the 
pure electroweak amplitude are induced by the leading 
four-quark operators ${\cal O}_{1,2}$. Thus the matrix 
elements $\langle s \ell^+ \ell^-| {\cal O}_{1,2}|b \rangle$ must be 
evaluated accurately, beyond the first non-trivial order. 
The corresponding Feynman diagrams are one-gluon-corrections to the basic one-loop 
diagrams with a charm quark contracted.  These correspond to 
two-loop integrals with three relevant scales: $m_b, m_c$ and $q^2$
(not simply two mass scales as in $B \to X_s \gamma$).
So far these diagrams have been analyzed only in Ref.~\cite{Asa1},
by means of a mass and momentum double expansion method.
By its very nature of expansion, the method of Ref.~\cite{Asa1}
is not applicable when the dilepton mass approach the $c\bar c$ threshold  
($q^2\sim 4m_c^2$) or becomes larger. We have applied a semi-numerical method, 
which first converts analytically all  two-loop integrals 
into a standard set of integrals and then performs a rapid numerical integration over a 
set of Feynman parameters.  This method is very accurate and works over any physical 
kinematical range, thus we are able to predict the partial decay rate not only in the 
low-mass window (as in  Ref.~\cite{Asa1}), but also above the  $c\bar c$ threshold.
As a result, our present work provides both a detailed and accurate
check of the results of Asatrian et al.~in the low mass window
and, at the same time, allow us to present the first NNLL results for 
the high-mass window.

\medskip

The NNLL perturbative calculations will bring predictions of 
 higher precision only if we can have the same level of confidence when estimating 
the non-perturbative effects.  These are divided into hadronization of the external 
$b$ and $s$ quarks, and the residual long-distance effects due to the tails 
of the narrow  $c \bar c $ resonances.
In both cases the corrections can be analyzed using appropriate heavy-mass expansions
and the resulting uncertainties will be shown to be under reasonable control
in both windows. In the high-mass range non-perturbative  uncertainties
turns out to be the dominant source of theoretical errors. However, 
as we shall show, a considerable reduction of this uncertainty can 
be expected in the near future with a better knowledge of 
universal non-perturbative parameters
from other processes.

\medskip

The experimental situation regarding the inclusive decay $B \rightarrow X_s
\ell^+ \ell^-$ is as follows: BELLE and BABAR have already obtained 
clear evidences  ($\approx 5 \sigma$) of this transition, quoting two 
measurements which are compatible with each other and 
with the SM expectation \cite{Bellebsll,Babarbsll}. 
Both results are based  on a semi-inclusive analysis: 
the hadronic system $X_s$ is reconstructed from a kaon 
plus $0$ to $4$ pions (at most one $\pi^0$). 
The signal characteristics 
is determined by modeling the invariant mass $M_{X_s}$ spectrum 
using the phenomenological model first proposed in \cite{ACCMM}. 
The reconstruction efficiencies of the signal are determined by the
MC samples based on this model, which leads to a large fraction 
of the present systematic uncertainty. 
The overall uncertainty of these first measurements of the inclusive decay
rate is still at the $30 \%$ level, but 
substantial improvements can be expect in the near future. 

\medskip

The plan of this paper is as follows:  in the next section we shall present  
a more thorough discussion of the effective-theory approach to inclusive FCNC
$b$ decays, briefly reviewing what is already known in the literature.
In Section 3, we shall describe in more detail the method we have used to calculate 
the two-loop matrix elements $\langle s \ell^+ \ell^-| {\cal O}_{1,\ 2} | b \rangle$.  
We shall give plots of these matrix elements and compare them with those obtained 
by Asatrian et al.: besides a very good agreement within 
the low-$q^2$ regime, they also show how and when the expansion method 
fails as we approach the $c\bar c$ threshold; our new results on the necessary 
one-loop counterterms are also presented. In Section 4 we present our new 
results  for the matrix element of ${\cal O}_8$ over the whole dilepton invariant mass spectrum.  
Section  5 is reserved for a concise analysis  of the non-perturbative effects
which also includes $1/m_b^3$ contributions.
.In Section 6 we finally present our phenomenological analysis: 
we quote results 
for the integrated rates over the two disjoint dilepton mass spectrum windows,
for the integrated lepton forward-backward asymmetry in the high mass region
and for the position zero of the asymmetry. A proper assignment of the theoretical 
uncertainties associated to these results will be presented.
A few appendices have been prepared to facilitate the reading 
of the main text.

\setcounter{equation}{0}
\section{Theoretical framework}

\subsection{Effective field theory}

Within inclusive $B$ decays, such as $B\to X_s \gamma $ and 
$B\to X_s \ell^+ \ell^-$, short-distance 
QCD corrections are sizable and comparable to  the pure electroweak 
contributions.  They stem from the  evolution of the system from a large scale
$M_{{\rm heavy}}=O(M_W)$, where the weak interaction acts, to the decay energy $\sim m_b$,
resulting in large logarithms of the form $\alpha_s^n(m_b)\,\log^m(m_b/M_{\rm heavy})$,
where $m \le n$ (with $n=0,1,2,...$).
The most suitable framework for their necessary resummations 
is an effective low-energy 
theory with five quarks, obtained by integrating out the
heavy degrees of freedom. 
Renormalization-group (RG) 
techniques are used to organize  the resummation of the series  in 
leading logarithms (LL), next-to-leading logarithms (NLL), and so on:

\be
\alpha_s^n(m_b) \,  \log^n(m_b/M)\quad [\mbox{LL}], \qquad  
\a_s^{n+1}(m_b) \, \log^n (m_b/M)\quad [\mbox{NLL}]~,\,\,  \ldots
\label{LLseries}
\ee
The effective five-quark low-energy Hamiltonian 
relevant to the partonic process 
$b \rightarrow s \ell^+\ell^-$ can be written as 
\begin{equation}
{\cal H}_{eff} = - \frac{4 G_{F}}{\sqrt{2}} V_{ts}^*V_{tb} \, 
\sum  {C_{i}(\mu, M_{\rm heavy})}\,\, \, {\cal O}_i(\mu) \, ,
\label{eq:effH}
\end{equation}
where 
\begin{equation}
\begin{array}{rlrl}
{\cal O}_{1} ~= &\!\!
(\bar{s} \gamma_\mu T^a P_L c)\,  (\bar{c} \gamma^\mu T_a P_L b)\,, & 
{\cal O}_{2} ~= &\!\!
(\bar{s} \gamma_\mu P_L c)\,  (\bar{c} \gamma^\mu P_L b)\,,  \\[1.2ex]
{\cal O}_{3} ~= &\!\!
(\bar{s} \gamma_\mu P_L b) \sum_q (\bar{q} \gamma^\mu q)\,,   &  
{\cal O}_{4} ~= &\!\!        
(\bar{s} \gamma_\mu T^a P_L b) \sum_q (\bar{q} \gamma^\mu T_a q)\,, \\[1.2ex]
{\cal O}_{5} ~= &\!\! 
(\bar{s} \gamma_\mu \gamma_\nu \gamma_\rho P_L b) 
 \sum_q (\bar{q} \gamma^\mu \gamma^\nu \gamma^\rho q)\,, &
{\cal O}_{6} ~= &\!\! 
(\bar{s} \gamma_\mu \gamma_\nu \gamma_\rho T^a P_L b) 
 \sum_q (\bar{q} \gamma^\mu \gamma^\nu \gamma^\rho T_a q)\,, \\[2.0ex]
\widetilde{{\cal O}}_{7}   ~= &\!\!     
  \displaystyle{\frac{e}{16\pi^2}} \, \mbMS(\mu) \,
 (\bar{s} \sigma^{\mu\nu} P_R b) \, F_{\mu\nu}\,,    &
\widetilde{{\cal O}}_{8}   ~= &\!\!  
  \displaystyle{\frac{g_s}{16\pi^2}} \, \mbMS(\mu) \,
 (\bar{s} \sigma^{\mu\nu} T^a P_R b)
     \, G^a_{\mu\nu}\,,        \\[2.0ex]                       
\widetilde{{\cal O}}_{9}   ~= &\!\!         
  \displaystyle{\frac{e^2}{16\pi^2}} \,
 (\bar{s} \gamma_\mu  P_L b)\, (\bar{\ell} \gamma^\mu \ell)\,, &
\widetilde{{\cal O}}_{10}  ~= &\!\! 
  \displaystyle{\frac{e^2}{16\pi^2}} \,
 (\bar{s} \gamma_\mu P_L b)\,  (\bar{\ell} \gamma^\mu \gamma_5 \ell)
\end{array}
\label{heffll}                                        
\end{equation}
define the complete set of relevant dimension-6 
operators; $C_{i}(\mu, M_{\rm heavy})$ are
the corresponding Wilson coefficients.
As the heavy fields are integrated out, the top-,
$W$-, and $Z$-mass dependence is contained in the 
initial conditions of these Wilson coefficients, 
determined by a matching procedure between the full and 
the effective theory at the high scale (Step~1). By means of RG equations, 
the $C_{i}(\mu, M_{\rm heavy})$ are then evolved to the low
scale (Step~2). Finally, the QCD corrections to the matrix 
elements of the operators are evaluated at the low scale (Step~3). 

\medskip 

Compared with the effective Hamiltonian relevant to 
$b \to s \gamma$, Eq.~(\ref{heffll}) 
contains additional operators $\widetilde{{\cal O}}_{9}$ and  
$\widetilde{{\cal O}}_{10}$ which are of order $\alpha_{em}$.
The first large logarithm 
of the form $\log(m_b/M_W)$  already appears without 
gluons, because the operator ${\cal O}_2$ mixes into $\widetilde{{\cal O}}_9$ 
at one loop. It is then convenient 
to  redefine the  magnetic, chromomagnetic and lepton-pair operators 
as follows~\cite{MM,BurasMuenz}:
\begin{equation}
\label{reshuffle}
{\cal O}_i = \frac{16\pi^2}{g_s^2} \widetilde{{\cal O}}_i~, 
\quad C_i = \frac{g_s^2}{(4\pi)^2} \widetilde{C}_i~, \quad \quad (i=7,...,10). 
\end{equation}
With this redefinition, one can follow
the three 
calculational steps discussed above \cite{MM,BurasMuenz}, 
 as in 
$b \to s \gamma$. 
In particular,  after the reshufflings in (\ref{reshuffle})
the one-loop mixing of the operator ${\cal O}_2$ 
with  ${\cal O}_9$ appears 
formally at order $\alpha_s$.
At NNLL precision, one should  in principle 
take into account the QCD two-loop corrections 
to $\langle {\cal O}_9 \rangle$, the QCD one-loop corrections to
$\langle {\cal O}_7\rangle $ and  $\langle{\cal O}_{10}\rangle$, 
and the QCD corrections to the electroweak one-loop matrix elements 
of the four-quark operators. 

\subsection{Present status of the calculation}

Regarding the present status of these perturbative 
contributions to decay rate and 
FB asymmetry of $B \to X_s \ell^+\ell^-$ (for a recent review see 
\cite{Review}), 
we note that the 
complete NLL contributions to the decay amplitude
can be found in \cite{MM,BurasMuenz}. 
Since the LL contribution to the rate turns out to be numerically 
rather small, NLL terms should be regarded an $O(1)$ correction 
to this observable. On the other hand, since a non-vanishing 
FB asymmetry is generated by the interference of vector  
($\sim {\cal O}_{7,9}$) 
and axial-vector ($\sim {\cal O}_{10}$) leptonic currents,
the LL amplitude leads to a vanishing result and therefore
NLL terms become by default 
the lowest non-trivial contribution to this observable.

\medskip 
For these reasons, a computation of NNLL terms
in $B \to X_s \ell^+\ell^-$ is needed 
if we are to aim for the same numerical accuracy as achieved 
by the NLL analysis of $B \to X_s \gamma$.
Before we proceed any further, we would like to acknowledge the
accumulated efforts by many groups, whose contributions greatly lessen
the toil.  Indeed, a large body of results for $b\to s \gamma$ 
can be taken over and used in the NNLL 
calculation of $B \rightarrow X_s \ell^+ \ell^-$  \cite{Adel,GH,Mikolaj,GHW,Burasnew}. The necessary 
{\it additional} calculations, including the two-loop 
corrections  presented in this  paper, 
have been cross-checked and our contribution here is a part of
that finalization process.  

\medskip 

To begin, the  
full computation of initial conditions 
to NNLL precision was  presented in Ref.~\cite{MisiakBobeth}. 
It removes the large matching scale uncertainty, which amounts 
to around $16 \%$  in the NLL approximation.

\medskip 
 
Part of the additional three-loop mixings (Step 2), 
which were not known from the 
$B \rightarrow X_s \gamma$ case, and which connect one 
of the four-quark operators 
${\cal O}_{1-6}$ to ${\cal O}_7$ and 
${\cal O}_9$, have been reported recently  
in \cite{Paolonew}: their effect leads to 
a $2 \%$ correction of the rate. 
The NNLL intra-mixings of the  four-quark operators 
are still missing but are expected to yield an even smaller contribution.  
Since the FB asymmetry does not receive contributions from the term
proportional to $|\langle \cO_9 \rangle|^2$, 
these mixings terms are not needed for a 
NNLL analysis of this observable.

\medskip 

In Step 3 the most important NNLL contributions come from the
two-loop matrix elements of the four-quark operators 
${\cal O}_1$  and  ${\cal O}_2$. They were calculated 
in \cite{Asa1}, using Mellin-Barnes techniques 
similar to the ones originally used in the corresponding 
$B \rightarrow X_s \gamma$ calculation \cite{GHW}.  
These lead to a double expansion in $m_c/m_b$ and $q^2/m_b^2$,
where $q^2$ is the dilepton mass squared. Thus, the results 
in \cite{Asa1} are only valid below the $c\bar{c}$ threshold within the 
dilepton mass spectrum. The calculation of the NNLL two-loop
matrix elements advocated by us in this paper is based on a semi-numerical method.
Since its validity spans over the whole dilepton mass spectrum, our work
not only gives an independent check of the calculation in \cite{Asa1}
within the 
low $q^2$ window, but stands on its own merits to give results 
above the $c \bar c$ threshold. 
A complete NNLL calculation also requires the one-loop matrix element 
of the operator ${\cal O}_8$. This was  done for 
small dilepton mass in \cite{Asa1}. We  now 
present results that  are valid
for arbitrary dilepton mass, to be reported below. 
In \cite{Asa1} it was shown  that within  the 
integrated low-dilepton mass spectrum 
these NNLL  matrix element contributions reduce the 
perturbative uncertainty (due 
the low-scale dependence) 
from $\pm 13\%$ down to $\pm 6.5\%$ and also the central value
is changed significantly, $\sim 14 \%$. We shall report our findings 
for both windows in due course. 
The  NNLL bremsstrahlung contributions and the corresponding IR virtual
one-loop corrections have also been calculated 
for the dilepton mass spectrum (symmetric part) in 
\cite{Adrian1,Asa1,Asa2} and for the FB asymmetry in \cite{Adrian1,Asa3,Asa4}. 
Their impact is significant; the zero of the FB asymmetry 
for example gets shifted by $11 \%$ by these contributions.

\medskip 

For completeness, we note that within Step 3 a few  
pieces are still missing but their effects are estimated not to exceed $2 \%$.
In particular, a complete NNLL calculation
of the $B \rightarrow X_s \ell^+\ell^-$ rate would require also 
the evaluation of the two-loop matrix element 
of the operator ${\cal O}_9$: its influence on the dilepton mass spectrum 
is expected to be  very small, because it gets multiplied by a 
small leading Wilson coefficient $C_9^{(0)}$. 
In any event, similar to the missing entries  of 
the anomalous-dimension matrix, this (scale-independent)
contribution does not enter into the FB asymmetry at NNLL accuracy.
The list of missing contributions includes also a calculation of 
two-loop matrix  elements of the four-quark (penguin) 
operators ${\cal O}_{3,4,5,6}$. However, their net NNLL effect  
is most likely to  be strongly
suppressed by their small Wilson coefficients. This expectation
is substantiated by the corresponding contributions to 
the $B \rightarrow X_s \gamma$ 
decay  branching ratio. The latter are  shown to have  an effect  
below $1 \%$ \cite{Misiak2002}. 

\medskip 

In conclusion, the QCD NNLL calculation of the FB asymmetry is now fully
complete, while the one for the dilepton mass spectrum distribution is 
on the verge of being completed.  Our present work is a part of that endeavor. 
All other missing pieces can be estimated to be smaller than $2 \%$. 
At this level of accuracy, other subleading effects  
may turn out to be more important.  In particular, 
the uncertainties regarding input parameters
should deserve a lot of  attention, and   
further studies regarding higher-order
electromagnetic effects are  necessary.  
 
\setcounter{equation}{0}
\subsection{Basic expressions}
In this subsection, we want to recapitulate some expressions and definitions of the basic observables. 
We normalize all by the semileptonic
decay width in order to reduce the uncertainties due 
to bottom quark mass and CKM angles:

\begin{equation}
\Gamma[ b \to X_c e \bar{\nu}_e] = 
\frac{ G_F^2 \mbpole^5 }{ 192 \pi^3} |V_{cb}|^2 f(z) \kappa(z)~.
\end{equation}
Here $z=\mcpole^2/\mbpole^2$ ($\mqpole$ denote pole quark masses), 
$f(z) = 1 - 8 z + 8 z^3 - z^4 - 12 z^2 \ln z$
is the phase-space factor and
\begin{equation}
\kappa(z) = 1 - \frac{2 \alpha_s(m_b)}{3 \pi} \frac{h(z)}{f(z)}
\label{semi}
\end{equation}
takes into account QCD corrections.  
The function $h(z)$ has been given analytically in \cite{NirNir}
and is explicitly displayed in one of our  appendices.
The normalized dilepton invariant mass spectrum is then defined as

\begin{equation}\label{decayamplitude}
R(s)=\frac{\frac{d}{d s}\Gamma(  B\to X_s\ell^+\ell^-)}{
\Gamma( B\to X_ce\bar{\nu})}~,
\end{equation}
where $s=(p_{\ell^+}+p_{\ell^-})^2/\mbpole^2$.
The other important observable is the forward--backward 
lepton asymmetry:
\begin{equation}\label{forwardbackward}
A_{\rm FB}(s)=\frac{1}{\Gamma( B\to X_ce\bar{\nu})}
  \int_{-1}^1 d\cos\theta_\ell ~
 \frac{d^2 \Gamma( B\to X_s \ell^+\ell^-)}{d s ~ d\cos\theta_\ell}
\mbox{sgn}(\cos\theta_\ell)~,
\end{equation}
where $\theta_\ell$ is the angle between the $\ell^+$ and $B$ momenta 
in the dilepton center-of-mass frame. It was  shown in \cite{Alineu}
that $A_{\rm FB}(s)$ is equivalent  to the energy asymmetry introduced in 
\cite{WylerMisiakCho}.
 
We also present here some useful formulae that will allow us  to 
systematically 
take into account all corrections to these two observables at the partonic level  
beyond the NLL approximation:

\bea
 R(s) = \frac{\alpha_{\rm em}^2}{4\pi^2}
\left|\frac{V_{tb}^* V_{ts}}{V_{cb}}\right|^2  && \!\!\!\!\!\!\!\! 
\frac{(1-s)^2}{f(z)\kappa(z)}    \left\{
   4 \left(1+\frac{2}{s}\right) |\Cnew_7(s)|^2   
   \left( 1+ \frac{\alpha_s}{\pi} \tau_{77}(s) \right) \right. \nonumber \\
&&  +(1+2s) \left[|\Cnew_9(s)|^2+|\Cnew_{10}(s)|^2\right] 
   \left(1+ \frac{\alpha_s}{\pi} \tau_{99}(s) \right)  \nonumber \\
&& \left. + 12\, \Re\left[ \Cnew_7(s)  \Cnew_9(s)^* \right] \left(1+  
   \frac{\alpha_s}{\pi} \tau_{79}(s) \right) + \frac{\alpha_s}{\pi}\delta_R(s) \right\}~,
   \qquad \label{NNLLDIMS} \\ 
\nonumber \\
A_{\rm FB}(s) = - \frac{3\alpha_{\rm em}^2}{4\pi^2}
 \left|\frac{V_{tb}^* V_{ts}}{V_{cb}}\right|^2  && \!\!\!\!\!\!\!\! 
 \frac{ (1-s)^2}{f(z)\kappa(z)}   \left\{  
   s \, \Re\left[ \Cnew_{10}(s)^* \Cnew_9(s) \right] 
  \left(1 +  \frac{\alpha_s}{\pi} \tau_{910}(s)\right) \right.  \nonumber\\
&& \left. + 2 \, \Re\left[ \Cnew_{10}(s)^* \Cnew_7(s) \right]
  \left( 1 + \frac{\alpha_s}{\pi}\tau_{710}(s) \right) + \frac{\alpha_s}{\pi}\delta_{\rm FB}(s)
  \right\}~.
\label{NNLLAFB}
\eea

\noindent 
In these expressions, we have introduced 
a new set of effective coefficients, defined as 
\bea
  \Cnew_7(s) &=&  \left(1+\frac{\alpha_s}{\pi}\sigma_7 (s)\right) 
       \widetilde C_7^{\rm eff} 
        -\frac{\alpha_s}{4\,\pi} \left[ C_1^{(0)} F_1^{(7)}(s)+
        C_2^{(0)} F_2^{(7)}(s) 
       + \widetilde C^{\rm eff(0)}_8 F_8^{(7)}(s) \right]  \nonumber \\
  \Cnew_9(s)  &=&  \left(1+\frac{\alpha_s}{\pi} \sigma_9 (s) \right) 
        \widetilde C_9^{\rm eff}(s) 
       -\frac{\alpha_s}{4\,\pi} \left[ C_1^{(0)} F_1^{(9)}(s) 
        + C_2^{(0)} F_2^{(9)}(s)+ \widetilde C_8^{\rm eff (0)} F_8^{(9)}(s) \right]
    \nonumber \\
  \Cnew_{10}(s)  &=& \left( 1+\frac{\alpha_s}{\pi} 
        \sigma_{9} (s) \right) \widetilde C^{\rm eff}_{10}~.
\label{effmod}
\eea
The $\Cnew_i$  have the advantage of encoding all dominant 
matrix-element corrections, which lead to an explicit $s$ dependence
in all of them.

The functions $\sigma_i$ 
and  $\tau_i(s)$ were calculated in \cite{Adrian1,Asa3}; their 
analytical formulae are given in an appendix.
The terms $\sigma_i(s)$ 
take into account {\em universal} $O(\alpha_s)$ bremsstrahlung,  
and the corresponding infrared (IR) virtual  corrections proportional 
to the tree-level matrix 
elements of ${\cal O}_{7-10}$. The remaining (finite) non-universal 
bremsstrahlung  connected with these operators are embedded in rate and 
FB asymmetry through  $\tau_i(s)$.
The additional finite terms $\tau_i(s)$
are rather small, especially 
for large values of $s$ ($|\tau_i(s)|< 0.5$ for $s>0.3$). 
The   finite bremsstrahlung corrections, not related to 
${\cal O}_{7-10}\otimes {\cal O}_{7-10}$,
are encoded in $\delta_{R,{\rm FB}}(s)$ and are substantially smaller. 
A complete evaluation of  
$\delta_{R}(s)$ can be found in~\cite{Asa2}, where     
its effect is shown to be at the $O(1\%)$ level. The effect of  $\delta_{FB}(s)$
is shown to be below $1\%$~\cite{Asa4}. 
The coefficients $\widetilde C^{\rm eff}_{7-10}$, including the 
one-loop matrix-element contributions of ${\cal O}_{1-6}$, 
are defined in close analogy with those in Ref.~\cite{BurasMuenz} and  
are written down  
in an appendix as a function of the true Wilson coefficients $C_i$. 
Finally, explicit expressions for the latter, evolved down to the 
low-energy scale, can be found in \cite{MisiakBobeth}.

\medskip

The other explicit $O(\alpha_s)$ terms in (\ref{effmod}) are 
due to virtual corrections that are infrared-safe. In particular, 
the two-loop functions 
$F_{1,2}^{(7),(9)}$ and  
the one-loop functions $F_8^{(7),(9)}$ are the matrix elements of 
 ${\cal O}_{1,2}$  and ${\cal O}_8$, respectively, including first order
$\alpha_s$ virtual corrections. 
These functions have been computed in Ref.~\cite{Asa1} for small $s$.
As for arbitrary dilepton mass, these functions will be a part of the main
fare in this article and will be discussed at  great length 
for the first time in subsequent sections.

\subsection{Organization of the perturbative ordering}
\label{sect:ordering}

As mentioned earlier, the so-called LL order as conventionally
labeled is not well 
justified  numerically, since the formally leading $O(1/\alpha_s)$
term in $\Cnew_9$ is much closer to being an $O(1)$ term.  
For this reason,  it has been proposed in Ref.~\cite{Asa1} to use 
a different counting rule, where the $O(1/\alpha_s)$
term of $\Cnew_9$ is treated as $O(1)$. We also subscribe to 
this approach. Although not consistently 
extendable to higher orders, it is operationally well defined 
for the present status of the calculation. Within this 
approach, the three $\Cnew_i$ and the two observables 
[$R(s)$ and $A_{\rm FB}(s)$] are all  
homogeneous quantities, starting with an $O(1)$ term.
Then all  $\sigma_i$, $\tau_i$ and $\delta_i$ functions are 
required for a NNLL  analysis of 
both $R(s)$ and $A_{\rm FB}(s)$. On the other hand, the missing
two-loop matrix element of the operator ${\cal O}_9$ and the 
three-loop mixing of the four-quark operators into  
${\cal O}_9$ \cite{Paolonew}, which otherwise should be a NNLL
contributor to the dilepton mass spectrum, are now 
formally of higher order within this new counting.

\setcounter{equation}{0}
\section{Calculation of the two-loop matrix elements of ${\cal O}_1$,${\cal O}_2$} \label{calculation}
\subsection{Two-loop diagrams}
\label{twoloopdiagrams}
\begin{figure}
\begin{center}
a) \psfig{figure=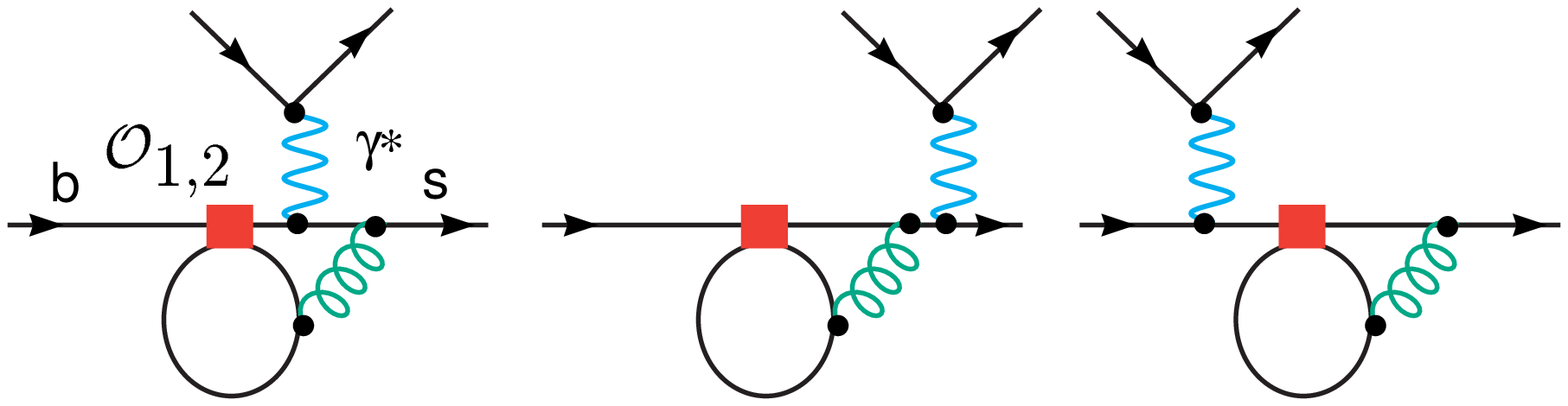,height=1.5in}\\
b) \psfig{figure=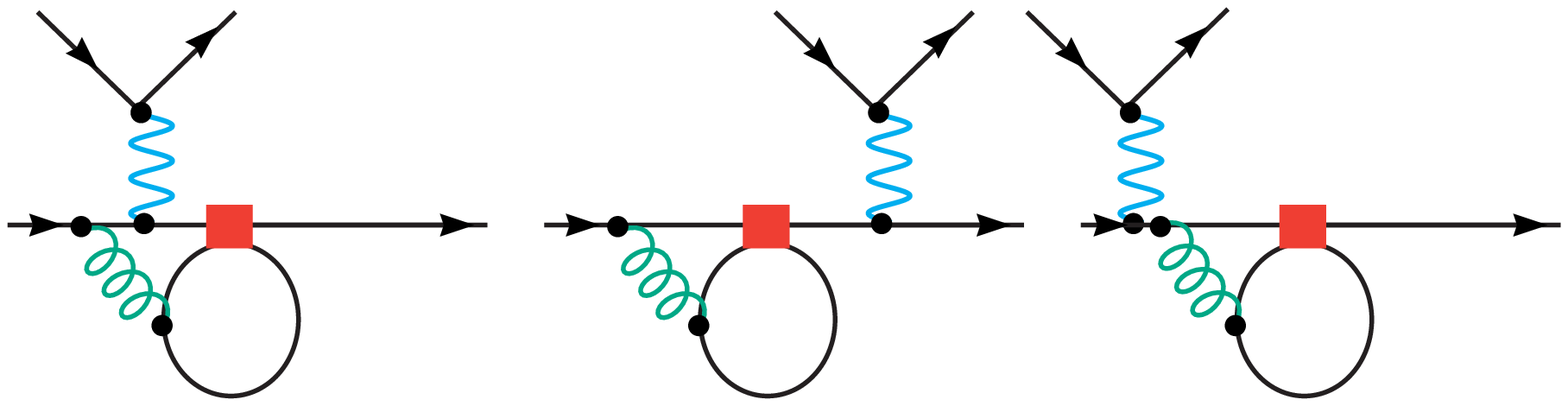,height=1.5in}\\
c) \psfig{figure=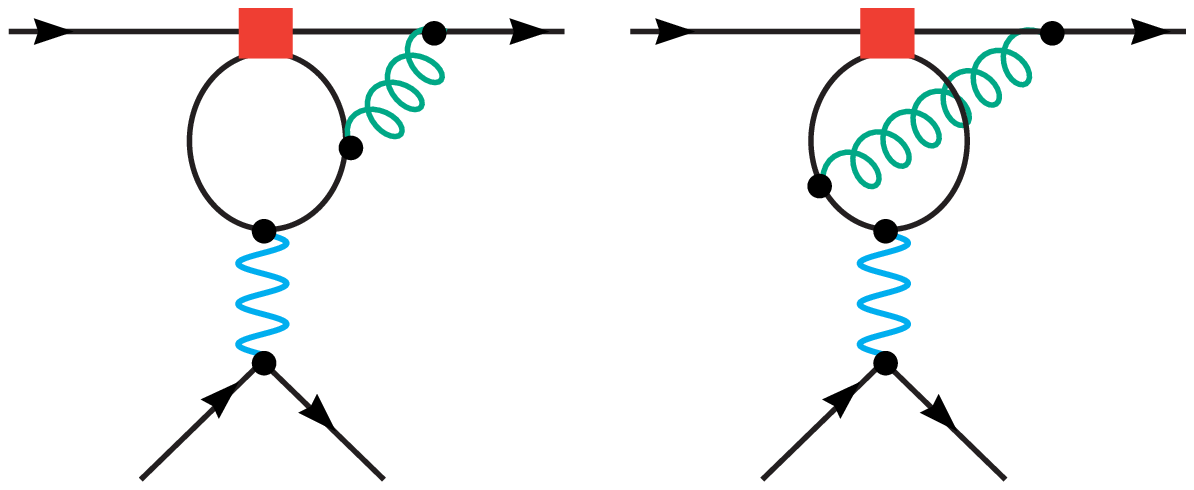,height=1.5in}\\
d) \psfig{figure=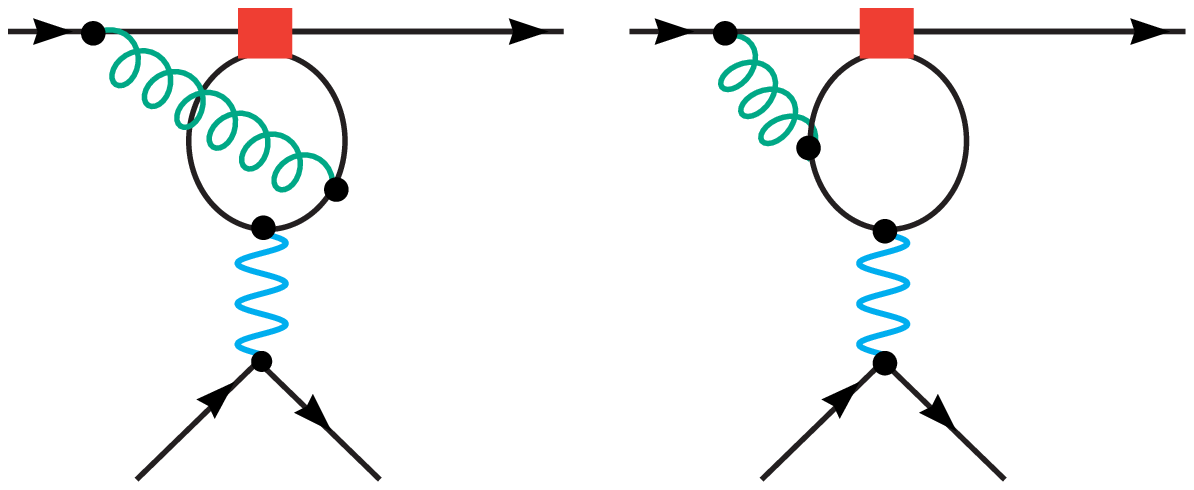,height=1.5in}\\
e) \psfig{figure=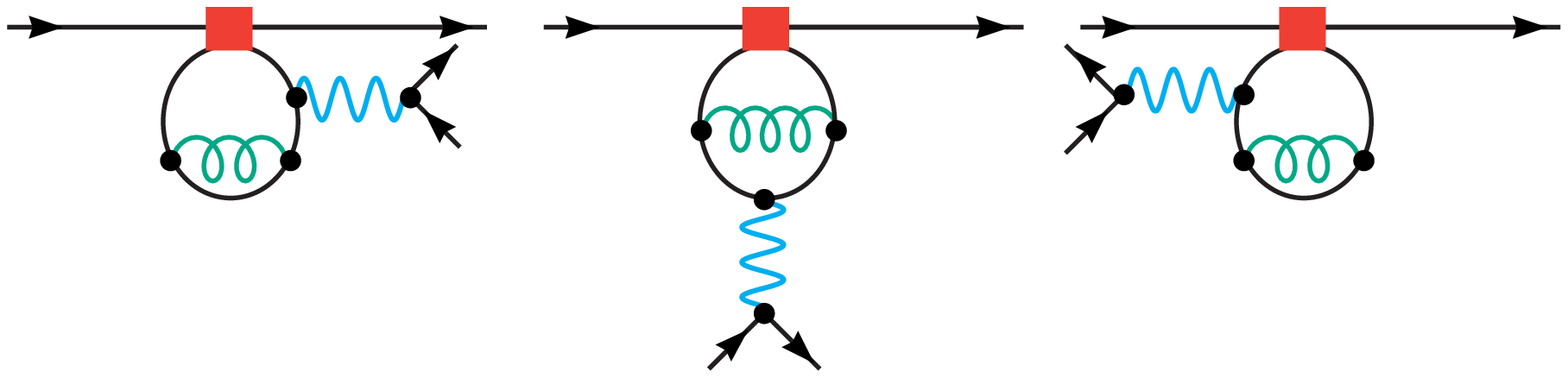,height=1.5in}
\end{center}
\caption{\footnotesize Two-loop Feynman diagrams relevant for the virtual QCD corrections
corresponding to the operators ${\cal O}_1$ and ${\cal O}_2$. 
They can be organized in five gauge invariant subsets.
\label{fig:radish}}
\end{figure}

\begin{figure}
\begin{center}
 \psfig{figure=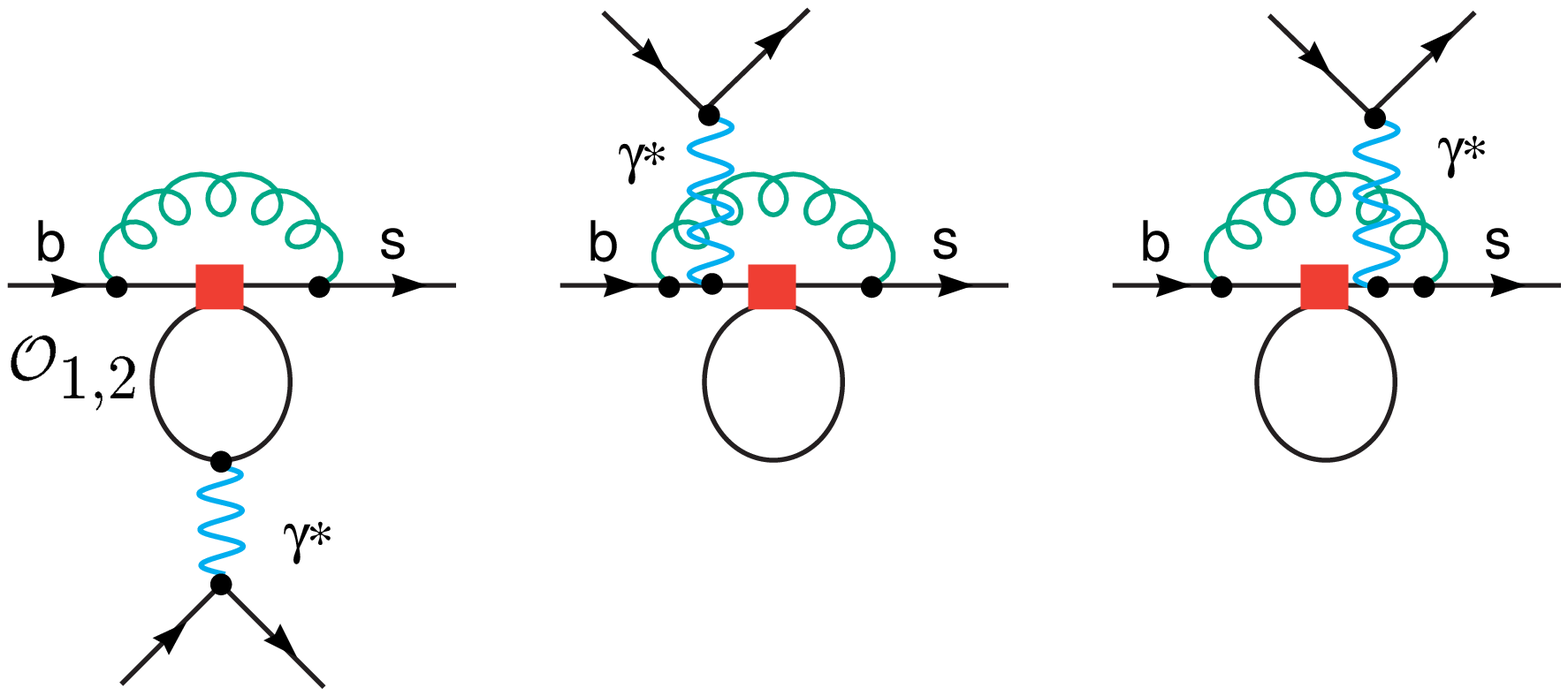,height=2.3in}
   \psfig{figure=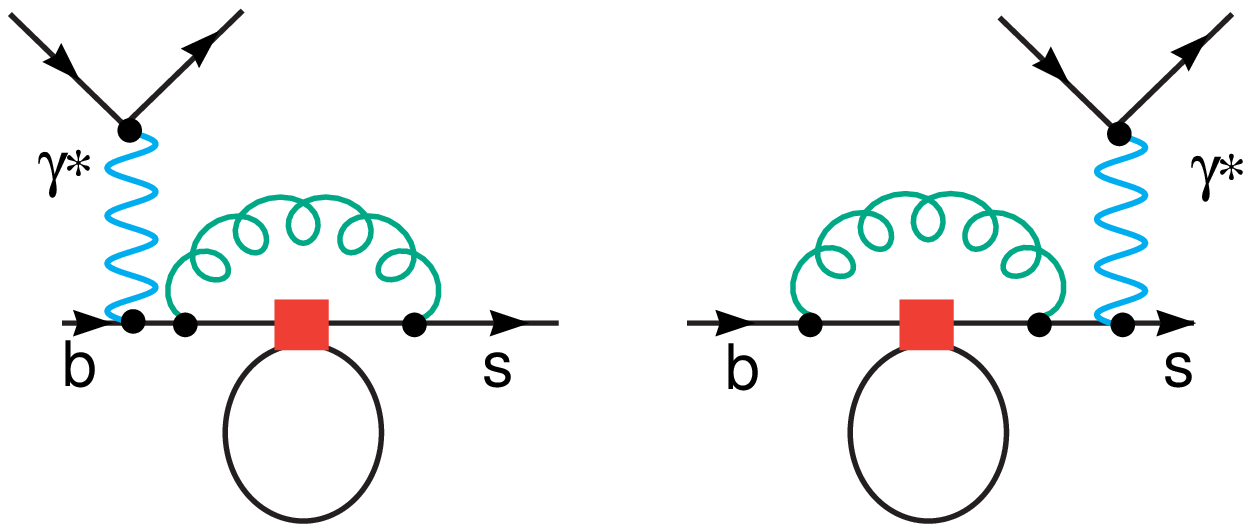,height=1.7in}
\end{center}
\caption{\footnotesize Two-loop Feynman diagrams relevant to the virtual
QCD corrections corresponding to the operators  ${\cal O}_1$ and ${\cal O}_2$, 
which can effectively be taken into account by a redefinition of the 
Wilson coefficients.
\label{fig:auto}}
\end{figure}

The relevant two-loop Feynman diagrams with 
${\cal O}_1$ and ${\cal O}_2$ insertions  
are shown in Fig.~\ref{fig:radish}. 
They will be organized into 
five gauge-invariant subsets. This is useful because gauge cancellations 
occur within each subgroup, and gauge invariance for each subset is a 
useful check of the calculation. 
We note that there is another subgroup of two-loop diagrams, 
shown in Fig.~\ref{fig:auto}. However, only the diagram 
where the virtual photon is attached to the charm loop is non-vanishing. 
The latter two-loop diagram factorizes into two one-loop
diagrams and can be included in the calculation 
of the virtual and bremsstrahlung contribution of the operator ${\cal O}_9$
due to their sharing the same Lorentz structure. 

This is most effectively done by the redefinition of the Wilson coefficient
(see \ref{effWilson} in the appendix). When calculating the corrections to the 
operator ${\cal O}_9$ using  these modified Wilson coefficients,  
the contributions of the diagrams given in \ref{fig:auto} are automatically
taken into account. 
We also note that by gauge invariance the QCD corrections to the matrix elements of the operators ${\cal O}_{1,2}$ can be written as  
\begin{equation}
\langle  s\ell^+\ell^- | {\cal O}_i\, | b \rangle = - \frac{\alpha_s}{4\pi} 
   \left(  F_i^{(9)} \langle \tilde{\cal O}_9 \rangle_{tree} +
    F_i^{(7)} \langle \tilde{\cal O}_7 \rangle_{tree}  \right). 
\label{gaugeinvariance}
\end{equation}

\subsection{Method}
\label{sec:method}

Within the $B \rightarrow X_s \gamma$ calculation
at NLL, the two-loop matrix elements of the four-quark operators  ${\cal O}_1$
and ${\cal O}_2$
for an on-shell photon were calculated in \cite{GHW} using Mellin--Barnes 
techniques. This calculation 
was extended in \cite{Asa1} to the case of an off-shell photon
with the help of a double Mellin--Barnes representation.
We are reminded that these matrix elements form an integral part of 
the NNLL analysis of the decay
$B \rightarrow X_s \ell^+\ell^-$.
The double expansion is in the dilepton mass $s = q^2/m_b2$ and 
the mass ratio $m^2_c/m^2_b$.
Thus, the validity of the analytical results given in \cite{Asa1} 
is restricted to small dilepton masses $s  < 0.25$,  because 
$c \bar c$ thresholds will  be crossed beyond that.

\medskip 

We follow here a different strategy to calculate these
two-loop matrix elements for {\it arbitrary} dilepton mass in our 
present NNLL work. 
\footnote{As mentioned in the introduction, the high-$s$ region above the 
$c \bar c$ resonances is experimentally also an important kinematic 
window since the efficiency is high there; thus, 
a comparable number of events will be collected there as in the 
low-$s$ region. However, as we will discuss in section \ref{nonperturbative}, 
one encounters larger non-perturbative corrections in this region.} 
  We use a universal method, which can evaluate 
any two-loop diagrams of general external kinematics and internal masses 
semi-numerically.  
For its implementation, the
diagrams are processed with a 
computer algebra program.  The aim of the various algebraic 
manipulations to follow
is to render the diagrams to a standard form,  which will be further 
integrated numerically in the second stage of the analysis.  
We used two independent versions written 
in FORM and in Schoonship,
which provide a powerful check on the algebra and consistency.
Let us describe the individual  steps of the calculation 
in more detail.  

\medskip

First, all two-loop diagrams are converted 
into sums of sun set type integrals and their mass derivatives:
\begin{equation}
 \int d^{n}p\,d^{n}q\, 
       \frac{p^{\mu_1} \ldots p^{\mu_i} q^{\mu_{i+1}} \ldots q^{\mu_j}}{
             ((p+k)^2+m_1^2)^{\alpha_1} 
             (q^2+m_2^2)^{\alpha_2}
             (r^2+m_3^2)^{\alpha_3}},
\end{equation}
where $r=p+q$, and $p$ and $q$ are the two independent internal momenta. 
This is done by the use of Feynman 
parameters $\{ X\}$ and appropriate shifts in the variables $p$ and $q$. 
\begin{figure}
\begin{center}
\psfig{figure=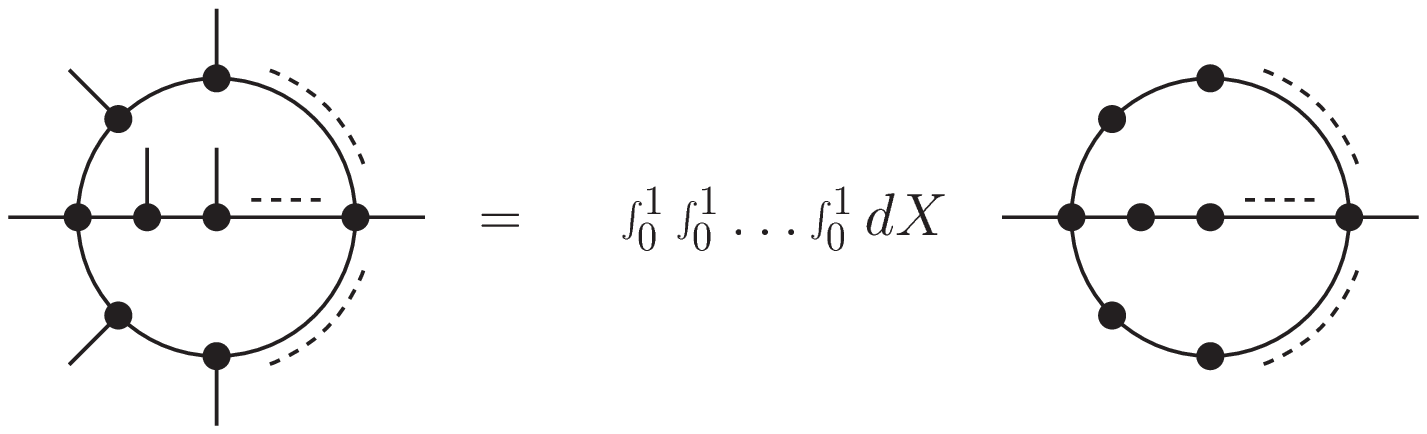,height=3cm} 
\end{center}
\caption{\footnotesize Expressing generic massive two-loop Feynman diagrams as 
integrals over sunset-type functions.
\label{fig:sun}}
\end{figure}
The effective masses $m_{1,2,3}^2$ and the 
effective momentum $k$ are 
polynomial functions of physical masses, external kinematics, and
Feynman parameters  associated with the diagrams (see Fig.~\ref{fig:sun}).
By using mass derivatives, there is 
in principle only one basic set arising 
from [$\alpha_1=\alpha_2=\alpha_3=1$] we need to know, although 
we shall instead use  [$\alpha_1=2, \alpha_2=\alpha_3=1$] as the basic set.
The reason for the latter  choice is that this set is less singular and 
it allows for neater 
integral representation of the corresponding scalar integrals, 
which makes them more suitable for numerical evaluation. 

\medskip

The second analytic step includes Lorentz decomposition of the tensor
structures and isolation of the scalar integrals, and use of 
differential recursion relations
to reduce the scalar functions to a set of ten master scalar functions.
The tensor reduction  
is done by decomposing the loop momenta $p$ and $q$ 
in the numerator into components 
parallel and orthogonal to the external momentum $k$, 
\begin{equation}
p_\perp^\mu=p^\mu-{p\cdot k\over k^2}k^\mu, \ \ 
q_\perp^\mu=q^\mu-{q\cdot k\over k^2}k^\mu .
\end{equation}
Tensor integrals with an odd number of transverse loop  momenta 
$p_\perp$ and $q_\perp$ vanish, while the even ones will be 
contracted basically with the metric tensor for further simplification.
The resulting scalar coefficients of the tensor decomposition 
are integrals of the following form:
\begin{equation} 
\tilde P^{ab}_{211}(m_1^2,m_2^2,m_3^2;k^2)=
\int d^np d^nq {(p\cdot k)^a (q\cdot k)^b 
\over [(p+k)2+m_1^2]^2(q^2+m_2^2)((p+q)^2+m_3^2)},
\label{scalarintegrals}
\end{equation} 
where $a+b \le3$ in renormalizable theories;
tensor integrals with more than three Lorentz 
indexes can be derived in a similar fashion but will not be needed.  
In general, one can prove \cite{method} that 
any two-loop diagram in renormalizable 
theories can be decomposed by this algorithm into an expression involving 
only ten scalar integrals, denoted by  ${\cal H}_i,\,\, i=1 \cdots 10$ in the 
following. They are linear combinations of integrals of the form 
(\ref{scalarintegrals}) with  $a+b \le3$. 
Let us reiterate that given enough computing power and 
up to possible infrared issues, which we shall mention below, 
the algorithm is applicable to any 
two-loop process.

\medskip

After this algebra step,  the  ten scalar integrals are integrated 
over $p$ and $q$.  An important feature is 
that the UV poles   
are isolated and one arrives at 
one-dimensional {\it finite}
integral
representations over a variable $0\le x\le 1$ of four 
elementary functions  of $m_{1,2,3}^2, k^2$  and $x$, 
which makes their numerical evaluation highly 
efficient and precise. 
As anticipated above, 
for this latter step our special choice 
$\alpha_1=2, \alpha_2=\alpha_3=1 $ as a basis of integrals 
is crucial; with this choice the ten basic scalar integrals ${\cal H}_i$
are logarithmically divergent in the ultraviolet and these UV divergences
are distinctly separate and manifestly exposed. 
The remaining finite parts, denoted 
by $h_i$, possess one-dimensional integral representations 
and will be displayed explicitly in an  appendix. 
The kernels of $h_i$ are, interestingly enough, 
moments in the variable $x$ of four elementary functions:
\begin{eqnarray} \label{elementary}
  \tilde{g} (m_1,m_2,m_3;k^2;x) & = &
     Sp(\frac{1}{1-y_1}) 
   + Sp(\frac{1}{1-y_2}) 
   + y_1 \log{\frac{y_1}{y_1-1}} 
   + y_2 \log{\frac{y_2}{y_2-1}} 
  \nonumber \\
  \tilde{f_1}(m_1,m_2,m_3;k^2;x) & = &
   \frac{1}{2}
   \left[
   - \frac{1-\mu^2}{\kappa^2}
   + y_1^2 \log{\frac{y_1}{y_1-1}} 
   + y_2^2 \log{\frac{y_2}{y_2-1}} 
   \right]
  \nonumber \\
  \tilde{f_2}(m_1,m_2,m_3;k^2;x) & = &
   \frac{1}{3}
   \left[
   - \frac{2}{\kappa^2} 
   - \frac{1-\mu^2}{2 \kappa^2}
   - \left( \frac{1-\mu^2}{\kappa^2} \right)^2
   + y_1^3 \log{\frac{y_1}{y_1-1}} 
   + y_2^3 \log{\frac{y_2}{y_2-1}} 
   \right]
  \nonumber \\
  \tilde{f_3}(m_1,m_2,m_3;k^2;x) & = &
   \frac{1}{4}
   \left[
   - \frac{4}{\kappa^2} 
   - \left( \frac{1}{3} + \frac{3}{\kappa^2}  \right) 
     \left( \frac{1-\mu^2}{\kappa^2} \right)
   - \frac{1}{2} \left( \frac{1-\mu^2}{\kappa^2} \right)^2
   - \left( \frac{1-\mu^2}{\kappa^2} \right)^3
   \right.
  \nonumber \\
 & &
   \left.
   \; \; \; \; \; \; \; \; \; \; \; \;
   \; \; \; \; \; \; \; \; \; \; \; \;
   \; \; \; \; \; \; \; \; \; \; \; \;
   + y_1^4 \log{\frac{y_1}{y_1-1}} 
   + y_2^4 \log{\frac{y_2}{y_2-1}} 
   \right]
   \; \; ,
\end{eqnarray}
%
\begin{eqnarray}
y_{1,2} =  \frac{1 + \kappa^{2} - \mu^{2}
                    \pm \sqrt{\Delta}}{2 \kappa^{2}}, \quad \quad 
\Delta  =  (1 + \kappa^{2} - \mu^{2})^{2} 
          + 4 \kappa^{2} \mu^{2} - 4 i \kappa^{2} \eta 
      \; \; , \nonumber
\end{eqnarray}
\begin{eqnarray}
   \mu^{2}  =   \frac{a x + b (1-x)}{x (1-x)}\, , \,\,\, 
         a  =   \frac{m_{2}^{2}}{m_{1}^{2}} \, , \; \; \; \;
         b \; = \; \frac{m_{3}^{2}}{m_{1}^{2}} \, , \; \; \; \;
\kappa^{2} \; = \; \frac{    k^{2}}{m_{1}^{2}} 
      \; \; .  \nonumber
\end{eqnarray}

\medskip

Once the analytical procedure  is done as described, each original 
Feynman diagram is  expressed as an integral over the set of Feynman 
parameters $\{ X\}$
introduced  earlier.  The integrand itself consists of a sum of 
the special functions $h_i$  (which are themselves 
one-dimensional integrals 
of the four elementary functions (\ref{elementary}))
and  possibly also of some trivial functions such as logarithms and 
rational functions of the 
kinematical invariants $m^2_{1,2,3},$ and $k^2$.  
Within our method, all these  integrations generally are
to be performed numerically.  We shall not dwell on the 
details here.  Suffice it to say that the analytic structure
of the functions $h_i$ is well understood, so that the 
integration paths $\{{\it C \}}$ 
can be moved into complex planes to effect better numerical convergence and, 
more importantly,  to 
yield amplitudes on the physical sheet.  Because we are interested in a 
high accuracy  and efficiency routine, we have used an adaptive deterministic 
integration algorithm.  
Such integration routines are very accurate, provided that the integrand 
is smooth enough and that the dimensionality of the integral is not too large. 
The integrand itself is, of course, an analytic function along any 
properly chosen  complex integration path of $\{{\it C \}}$, and 
therefore in order to preserve this smoothness, it is advantageous to optimize
the choice for a smooth 
integration path
in the numerical work as well.  One should be made aware that the integrals over Feynman 
parameters must be performed along a complex integration path that is 
consistent with the causality condition. This path is computed 
automatically by using spline functions such 
that both the path itself and its Jacobian are
smooth functions. Moreover, we note that,  in the problem at hand, we shall be dealing
with three-fold numerical integrations at most.

\medskip

We would like to mention that we perform minimal subtractions 
on all divergent subgraphs.  We have checked that the anomalous dimensions so 
obtained for the operator mixing matrix elements agree with what is known
 in the literature.  
This will be further elaborated in a subsequent section.  
We have explicitly checked that each group of diagrams is 
gauge-invariant and have been further
reassured by their numerical stability.

\medskip 

In the case of the calculation of the two-loop matrix elements 
of $B \rightarrow X_s \ell^+ \ell^-$ decay, we deal with 
three kinematic variables:
the charm mass, the dilepton invariant mass, and the subtraction scale. 
In order for the result to be usable for 
phenomenological studies,
in particular to be implementable into a Monte Carlo simulation, we need to 
cover this 
three-dimensional kinematic space. A real-time calculation of the two-loop 
matrix element
by numerical integration is far too slow to be implemented directly into a 
Monte Carlo
simulation of the experimental set-up.
With the present-day processors, 
an alternative of calculating in advance a comprehensive grid 
of integration points
that cover the whole three-dimensional kinematic range, 
storing them, and 
then interpolating between them,  seems to be the most efficient  
way because of the semi-numerical nature of our whole approach. 
All these considerations led us into writing a program that calculates the 
two-loop matrix elements efficiently and accurately,
and which is fast enough to be incorporated into a Monte Carlo simulation.
We selected a grid of $38 \times 3 \times 3$ integration points for 
both the electric 
and the magnetic components of the two-loop virtual corrections. 
Each of the 684 integration point yielding values for the form factors 
was calculated with a relative precision of $10^{-3}$. 
We used the CERN
Linux cluster to perform this calculation, and the CPU usage was 
approximately 3 days on 33 processors (mostly 850 MHz) running in parallel.

\medskip 

Finally, we would like to mention a caveat in the semi-numerical algorithm 
presented here.  In a general process, 
infrared singularities will most likely occur and will lead  
to infrared divergences in our  integral
representations. In such cases, it is more efficient to first 
separate out the 
infrared parts of the two-loop diagrams in an analytically manageable
form. 
Fortunately, for the process at hand all relevant diagrams (a)--(e) 
in Fig.~\ref{fig:radish} 
are infrared-finite; therefore the algorithm is most suitable for this 
specific application.

\subsection{Unrenormalized results}

In the following we show plots of our  results for the finite ($\epsilon^0$) 
parts of  the unrenormalized (naked)  two-loop Feynman diagrams of 
Fig.~\ref{fig:radish}, 
which give matrix elements to the  operators ${\cal O}_1$ and ${\cal O}_2$ 
within the $\overline{MS}$
scheme. The finite counterterm contributions are not included here but
will be discussed in the next subsection. 
Our main purpose is to compare our results with those of Ref.~\cite{Asa1}  
where Mellin--Barnes expansion techniques were used. 

\medskip 

\begin{figure}
\begin{center}
\psfig{figure=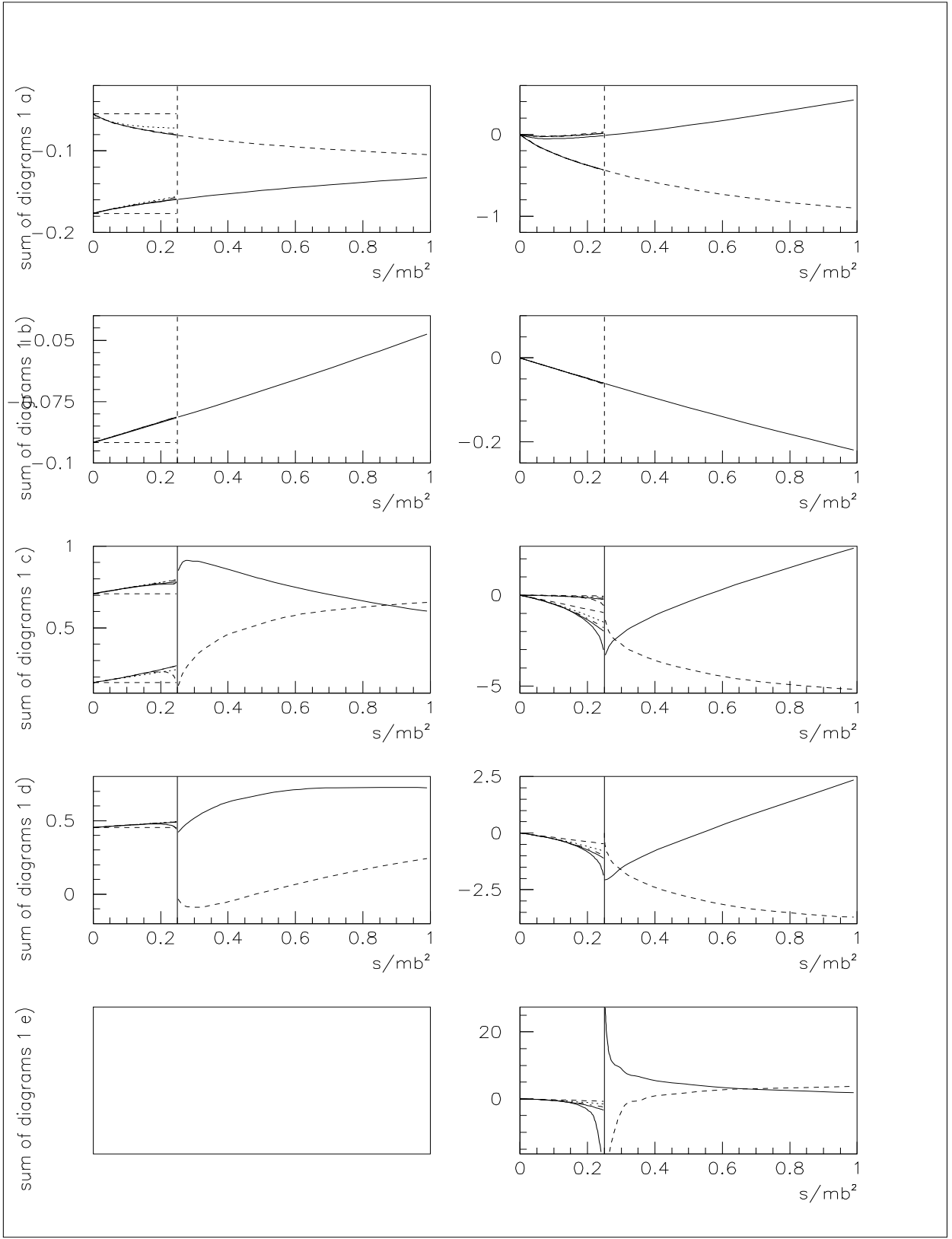,height=8.5in}
\end{center}
\caption{\footnotesize Plots of the UV-finite part of the
Feynman diagram subsets shown in Fig.~\ref{fig:radish}.
The two columns correspond 
to the electric and the
magnetic form factors; see (\ref{gaugeinvariance}). There is 
no electric contribution from the diagrams 1 e). 
 We plot the real (solid line) and the imaginary (dashed line) parts of 
our exact numerical integration result, 
along with successive approximations in the 
momentum expansion series
of the result of Asatrian et al. which is shown only below the $c\bar c$ 
threshold. For more details see main text. 
\label{fig:results}}
\end{figure}

Each plot in  Fig.~\ref{fig:results}  represents one of the five 
gauge-invariant diagram subsets given in Fig.~\ref{fig:radish}. 
We want to bring attention to  the complete range of the
dilepton mass spectrum $s = [0,1]$ in all the plots, which is a salient point 
of this discussion.  For each we plot the electric ($F^{(7)}$) and
the magnetic ($F^{(9)}$)  
contributions, defined in (\ref{gaugeinvariance}) 
separately in the left and right columns, except for the subset (e), 
which has  no contribution in the first column.
The calculation of Asatrian {\it et al.} is shown as 
successive approximations, showing that it converges toward our result.
Their results are valid only under the $c \bar c$ threshold, which we
demarcate by  vertical lines in the figures. Generally, 
the real part of our results is given by 
a solid line and the imaginary
part by  a dashed line. All the other lines are successive approximations 
in the momentum expansion series of Asatrian {\it et al.}.

\medskip 
  
\begin{figure}
\begin{center}
\psfig{figure=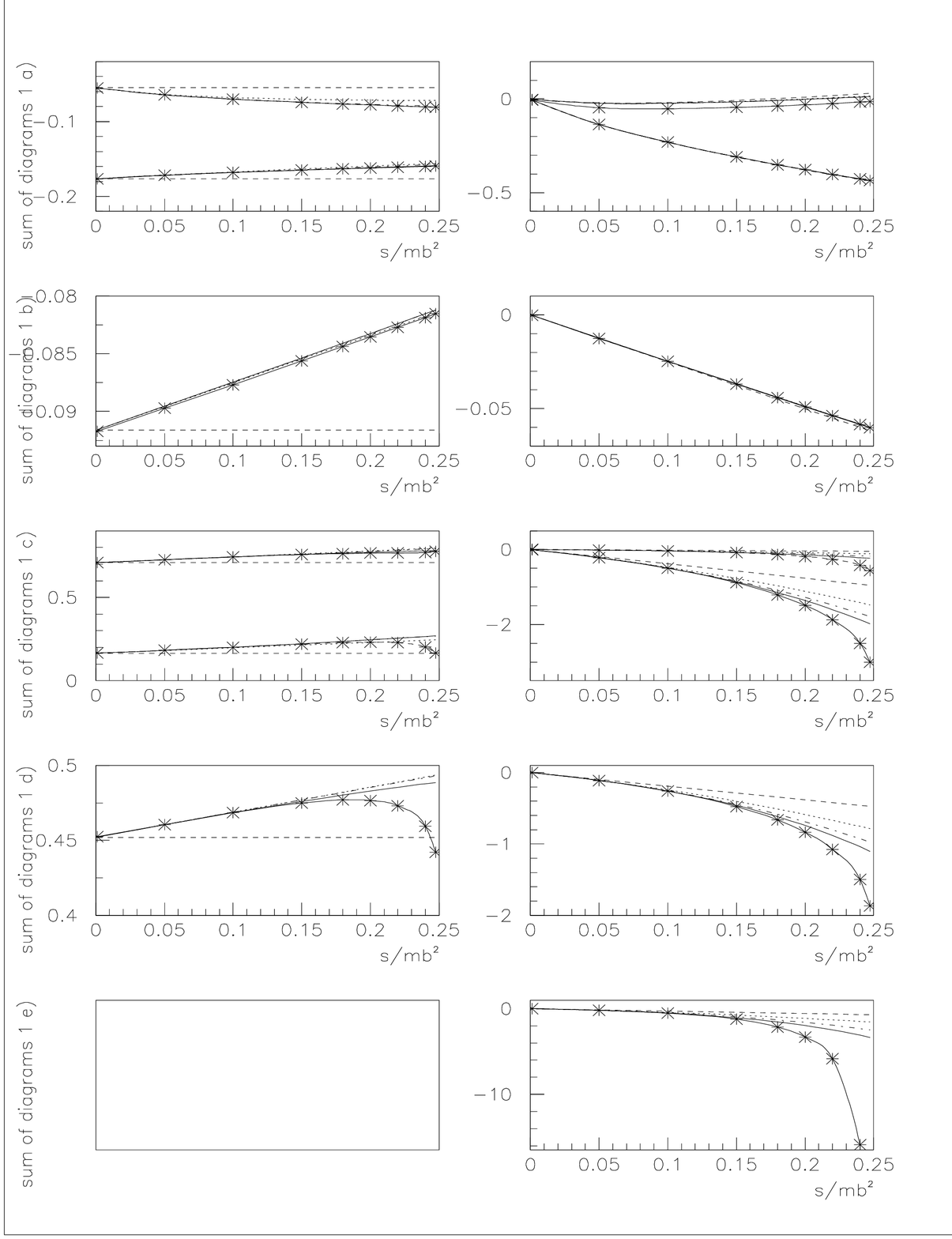,height=8.5in}
\end{center}
\caption{\footnotesize
Convergence pattern of the momentum expansion solution of 
Asatrian {\em et al.}
toward our numerical integration solution. The stars denote actual 
integration points, and the 
solid line connecting them is an interpolation. For more details 
see main text. 
\label{results2}}
\end{figure}

In  Fig.~\ref{results2} we zoom in to magnify the way the momentum expansion 
converges toward  our exact numerical solution within the low-$s$ region.
Here the stars are the actual points obtained from our numerical output,
which are joined by the continuous lines obtained from the 
interpolation program.

\medskip 

The comparison can be summarized as follows: 
there is good agreement between our results for each diagram set and the 
double expansion of Asatrian {\it et al.}  in the region below
the $c \bar c$ threshold. 
This agreement within the low-$s$ region 
provides a strong confirmation of our numerical method.
One notices that, as a general rule, the less singular the threshold 
behaviour of the diagram is,
the better the momentum expansion converges toward our exact numerical 
result. 
For subsets (a) and (b) the expansion converges best
because of the lack  of a nontrivial threshold. 
For (c) and (d) the threshold is
relatively mild, and the convergence is intermediary. For the gauge-invariant
 subset (e),  the threshold
is quite singular and thus we notice the poorest convergence of the 
momentum expansion. 
This singular behaviour is mostly due to the charm 
self-energy-type diagrams, as evidenced by a
much milder disagreement of the two methods after the mass counterterm is added.  
This also makes the agreement between our final physical result 
and the momentum expansion result better than what can 
be inferred from Fig.~\ref{results2} alone. 
The actual agreement of the two calculations 
within the low-$s$ 
region is compatible with the error of our numerical integration accuracy.
We may find it interesting that, for the gauge-invariant subsets (a) and (b) 
the expanded results of \cite{Asa1}  
are  actually valid beyond the $c \bar c$ threshold, perhaps 
because these diagrams have  no threshold cuts in the way.

\medskip

Before leaving this subsection, we want to remark that by our method we have 
reproduced numerically the values given by the analytical results on the 
two-loop matrix elements of ${\cal O}_{1,2}$  
in the  $B \rightarrow X_s \gamma$ mode presented in \cite{GHW},
with an accuracy well below  $1 \%$.

\subsection{Counterterm contributions} 

The counterterms  to the matrix elements of the operators 
${\cal O}_1$ and ${\cal O}_2$ will serve two purposes.
They give  renormalization 
to the QCD parameters and they account for operator renormalization
due to  mixing. Although the counterterms expanded 
in the variable $s$ can be found in \cite{Asa1},  
for us we need to generalize  to 
arbitrary dilepton mass.  Before giving our new final analytical results, 
we commence with a
short discussion of the various bits and pieces that must
be put together. 
We follow  the conventions in \cite{Asa1}, including the convention
$d=4-2\, \epsilon$ and their convention regarding the evenescent
operators.  The anomalous dimensions describing the operator mixing are 
known (see \cite{MisiakBobeth,Asa1,Paolonew}) and are given 
in the appendix. We take this opportunity to point out a very 
useful feature of  
our semi-numerical method in that the UV divergences are 
separated and given analytically through the whole calculation.
Because of this, we were able to  check 
the two-loop mixing results explicitly.

\medskip 

The list of non-trivial counterterm contributions 
to the functions $F_i^{(9)}$ 
and $F_i^{(7)}$ in (\ref{gaugeinvariance}), which are  
add to  those from the naked diagrams given in Fig.~\ref{fig:radish}, 
is the following: 

\begin{itemize}

\item  The two operators ${\cal O}_1$ and ${\cal O}_2$, which 
non-trivially mix into the four-quark operators at the one-loop level, 
inducing additional counterterm 
contributions.   We denote them
by  $F_{i \rightarrow {4quark}}^{(7)}$ and $F_{i \rightarrow {4 quark}}^{(9)}$.
They are given by 
\begin{equation}
\label{countertermfour} 
\sum_{j} \frac{\alpha_s}{4\pi}  \frac{1}{\epsilon} 
 a_{ij}^{11} \langle s \ell^+ \ell^- |  {\cal O}_j | b \rangle_{one-loop}
= -(\frac{\alpha_s}{4\pi} ( F_{i \rightarrow {4 quark}}^{(7)}  
 \langle \tilde{\cal O}_7 \rangle_{tree} + F_{i \rightarrow {4 quark}}^{(9)} 
\langle \tilde{\cal O}_9 \rangle_{tree})),
\end{equation}
in which $j$ runs over all four-quark operators. This expression
instructs us to calculate 
the one-loop matrix elements of the four-quark operators 
up to the  $\epsilon^1$ terms for arbitrary dilepton mass.

\item 
The analogous one-loop mixing of the operators 
${\cal O}_1$ and ${\cal O}_2$ into ${\cal O}_9$. 
They are from the  two-loop diagrams 
shown in Fig.~\ref{fig:auto}. 

\item 
The one and two-loop mixing into ${\cal O}_9$, which is   
connected with the renormalization of the explicit 
coupling constant in the definition of
 ${\cal O}_9$. They appear in the 
diagrams of Fig.~\ref{fig:radish}. This leads to an additional contribution
to $F_i^{(9)}$ (but not  $F_i^{(7)}$)     in (\ref{gaugeinvariance}),
which we call 
 $F_{i \rightarrow {9}}^{(9)}$.   
It is given by 
 \begin{equation}
 F_{i  \rightarrow  9}^{(9)} = -(\frac{a_{i9}^{22}}{\epsilon^2} +
\frac{a_{i9}^{12}}{\epsilon}) -
 \frac{a_{i9}^{11} \, \beta_0}{\epsilon^2},
  \end{equation}
where we have used the coupling renormalization 
\begin{equation}
    Z_{g_s} = 1 - \frac{\alpha_s}{4\pi} \, \frac{\beta_0}{2} \,
\frac{1}{\epsilon} + O(\alpha_s^2) \quad \mbox{with}\,\,
\beta_0 = 11 - \frac{23}{3} 
\quad \mbox{for five 
active flavours}.
\end{equation}

\item 
Then, there are QCD mass counterterm contributions
from the 
renormalization of the charm mass within the matrix elements of
${\cal O}_1$ and ${\cal O}_2$, 
which we denote by $F_{i,m_c}^{(9)}$ and 
$F_{i, m_c}^{(7)}$. They are 
most easily given by 
replacing the charm mass $m_c$ by $Z_{m_c} \times m_c$  in the
one-loop matrix elements of ${\cal O}_1$ and ${\cal O}_2$. 
We use the pole mass renormalization in our calculation:
\begin{equation}
\label{fieldmb}
    Z_{m_c} = 1 - \frac{\alpha_s}{4\pi} \frac{4}{3} 
\left(\frac{m_b^2}{\mu^2}\right)^{- \epsilon}
        \left( \frac{3}{\epsilon} + 4 \right) +O(\alpha_s^2),
\end{equation}

\end{itemize}

In the following, we give
the complete finite ($\epsilon^0$) counterterm contributions
from the two-loop diagrams in Fig.~\ref{fig:radish}.  
 We  write down our new final results for the 
relevant finite ($\epsilon^0$) parts for arbitrary dilepton mass:
\begin{equation}
F^{(k)}_{i,ct} =  F_{i  \rightarrow  9}^{(k)} +  
F_{i  \rightarrow  4quark}^{(k)} +  F_{i,m_c}^{(k)}, \quad k=7,9\, , i=1,2. 
\end{equation}
\begin{eqnarray}
 F^{(7)}_{1,ct}\,\,  |_{\epsilon^0}  &=& \frac{4}{81} l_{\mu} + \frac{4}{81} B(s)  \\
 F^{(7)}_{2,ct}\,\,  |_{\epsilon^0}  &=& -\frac{8}{27} l_{\mu} - \frac{8}{27} B(s) \\
&&  \qquad \no \\
 F^{(9)}_{1,ct}\,\,  |_{\epsilon^0}  &=& i\pi\left[ \frac{8}{243} l_s  
   + \frac{8}{243}  l_{\mu}  - \frac{16}{729} \right]
   + \frac{16}{729} l_s + \frac{704}{81} l_{c} 
   - \frac{3560}{2187}  - \frac{4}{243}  l_s^2 
   - \frac{8}{243} l_s l_{\mu}  \qquad
\no \\ &&  
    + \frac{16}{27}  l_{c} l_{\mu}   
    +\frac{256}{9}  C_x (s_c) l_{\mu}  
    +\frac{32}{9}   B_x (s_c)   l_{c} 
    +\frac{32}{9}  B_x (s_c)     l_{\mu}
    +\frac{256}{9} C_x (s_c)   l_{c} 
\no \\ &&  
    +\frac{256}{9} C_{xx} (s_c)   l_{c} 
    -\frac{256}{9}   C_{xx} (s_c)    l_{\mu}   
    +\frac{6328}{729}     l_{\mu}   
    -\frac{32}{9}   B_{xx}(s_c) l_{\mu}   
   - \frac{32}{9}   B_{xx}(s_c)  l_{c}  
\no \\ &&  
   +\frac{16}{81}   B_{xx}(s) 
   - \frac{16}{81}   B_x (s)  
   - \frac{8}{81}  B_{2x}(s)  
   +\frac{16}{9}   B_{2x}(s_c) 
   - \frac{640}{27}  B_{xx}(s_c)  
   + \frac{640}{27}  B_x (s_c)  
\no \\ &&  
   + \frac{16}{243}   \pi^2 
   + \frac{8}{81}  B_{2xx} (s) 
   - \frac{16}{9}   B_{2xx} (s_c)  
   + \frac{128}{27}  C_{xx}(s_c)  
   + \frac{64}{9}   C_{2x}(s_c)
\no \\ && 
   - \frac{64}{9} C_{2xx}(s_c)   
   - \frac{16}{81}  l_{\mu}   B_x (s)  
   + \frac{16}{81}   l_{\mu}  B_{xx}(s)  
   +\frac{ 64}{243}   l^2_{\mu} 
   -\frac{128}{27}   C_x (s_c) \\
&& \quad \no \\
 F_{2,ct}^{(9)}  \,\, |_{\epsilon^0}  &=& i\pi\left[
    \frac{32}{243} - \frac{16}{81} l_{\mu} -\frac{16}{81}  l_s \right]
   -\frac{32}{243}   l_s 
   + \frac{32}{27} l_{c}  + \frac{8}{81}   l_s^2   
   + \frac{16}{81} l_s  l_{\mu}   
   - \frac{16}{9}   l^2_{c}
   - \frac{32}{9}  l_{c} l_{\mu} 
\no \\ &&  
   + \frac{64}{3}  C_x (s_c)     l_{\mu}  
   - \frac{64}{3}  B_x (s_c)   l_{c}    
   - \frac{64}{3}   B_x (s_c)     l_{\mu}  
   +\frac{64}{3}   C_x (s_c)   l_{c}   
   - \frac{64}{3}  C_{xx}(s_c) l_{c}    
\no \\ &&          
   - \frac{64}{3}   C_{xx} (s_c)     l_{\mu}   
   +\frac{304}{243}   l_{\mu} 
   +\frac{ 64}{3}  B_{xx}(s_c)    l_{\mu} 
   + \frac{64}{3}  B_{xx}(s_c) l_{c}
   - \frac{32}{27}  B_{xx}(s)  
  + \frac{32}{27}   B_x (s)
\no \\ &&  
  +\frac{16}{27}   B_{2x}(s) 
  - \frac{32}{3}  B_{2x}(s_c)   
  +\frac{128}{9}   B_{xx}(s_c)  
  - \frac{128}{9}   B_x (s_c)
  - \frac{32}{81} \pi^2 
   - \frac{16}{27}    B_{2xx}(s) 
\no \\ &&  
   +\frac{32}{3} B_{2xx}(s_c)  
  + \frac{32}{9}   C_{xx}(s_c) 
  + \frac{16}{3} C_{2x}(s_c) 
   - \frac{16}{3}   C_{2xx} (s_c)  
   +  \frac{32}{27}  l_{\mu}    B_x(s)
\no \\ &&  
   - \frac{32}{27}    l_{\mu}  B_{xx}(s) 
  - \frac{128}{81}    l^2_{\mu}  
  - \frac{32}{9}   C_{x}(s_c)  
  - \frac{656}{729} 
\end{eqnarray}
The functions $B(a)$, $B_x(a)$, $B_{xx}(a)$, $B_2 (a)$, $B_{2x}(a)$, $B_{2xx}(a)$
and also the functions $C(a)$, $C_x (a)$, $C_{xx} (a)$, $C_2 (a)$, $C_{2x} (a)$, $C_{2xx} (a)$
are defined in the  appendix.  We have also introduced the variable
$s_c = q^2/m_c^2$ and we have defined 
$l_\mu=\ln(m_b^2/\mu^2)$,  $l_{c}=\ln(m_c^2/m_b^2)$, and $l_{s}=\ln(s)$.

We have explicitly checked that all the $\epsilon^{-2}$ and 
$\epsilon^{-1}$ terms coincide with the results given in \cite{Asa1}.

\section{Calculation of the $O(\alpha_s)$ virtual corrections to the\\
matrix element of ${\cal O}_8$}

\begin{figure}
\begin{center}
\psfig{figure=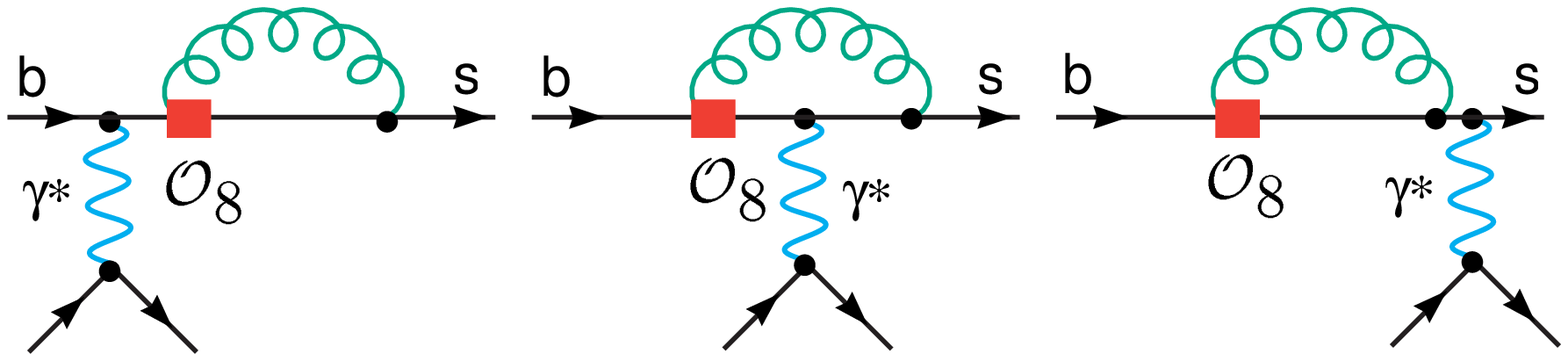,height=1.3in}
\psfig{figure=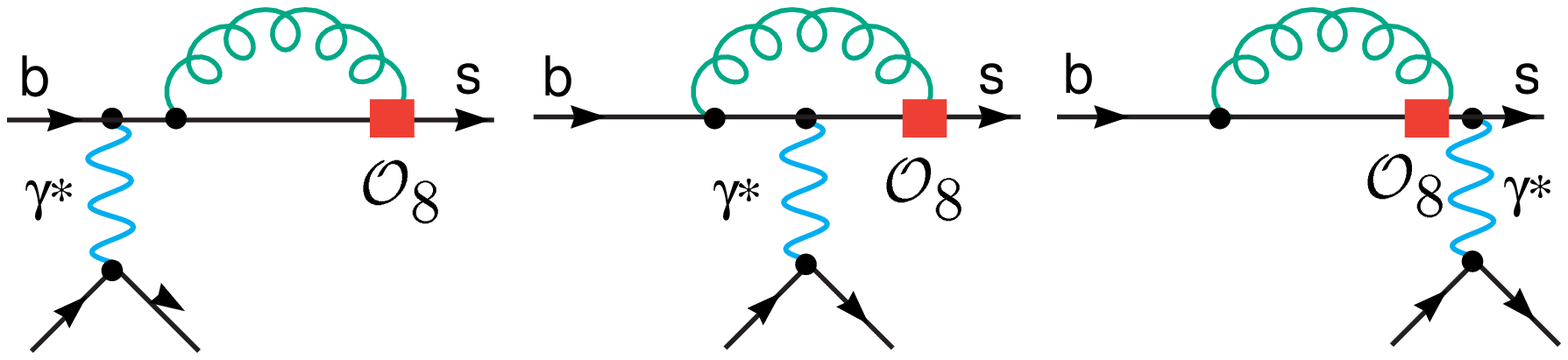,height=1.3in}
\end{center}
\caption{\footnotesize One-loop Feynman diagrams relevant to the virtual QCD corrections
corresponding to the operators ${\cal O}_8$.
\label{fig:08}} 
\end{figure}

We present here a short description of our new calculation 
of the matrix element of the operator ${\cal O}_8$ for general $s$. 
 Besides the contributions from the naked 
diagrams shown in Fig.~\ref{fig:08}, 
there is also a counterterm contribution due to the mixing of ${\cal O}_8$ 
into ${\cal O}_7$:  
\begin{equation}
\langle s\ell^+\ell^- |  \delta Z_{87} \, {\cal O}_8 | b \rangle, 
\quad \quad \delta Z_{87} =
- \frac{\alpha_s}{4\pi} \frac{16}{9\epsilon}\,. 
\end{equation}
The functions $F_8^{(7)}$ and $F_8^{(9)}$ in (\ref{effmod})
are then defined by the renormalized matrix element of ${\cal O}_8$:
\begin{equation}
   \langle s \ell^+ \ell^- | C_8 {\cal O}_8  | b \rangle =
    {C}_8^{(1)} ( -\frac{\alpha_s}{4 \pi} )
    ( F_8^{(9)} \langle \tilde{\cal O}_9 \rangle_{tree} +
        F_8^{(7)} \langle \tilde{\cal O}_7 \rangle_{tree} ),
\end{equation}

Keeping the full $s$ dependence, we find 

\bea
F_{8}^{(7)} &=&  - \frac{32}{9} \ln\left(\frac{\mu}{m_b}\right) 
 - \frac{4}{27} \pi^{2} - \frac{4}{9}  -\frac{8}{9} i \pi  
 + \frac{8}{9} \frac{ s\ln(s) }{s - 1}  \no \\
&& +  \frac{4}{9} \int_{0}^{1}dx  \int _{0}^{1 - x}dy~\left[ 
(7 - 3x - 6y) \ln\left( 1- s \frac{xy}{x^2 + xy + y} \right)  \right.
  \no \\
&& \qquad\qquad \left. - \frac{ x^3 + 3x^2y + 2xy^2 - xy - s(x^2y + 2xy^2 - 3xy)}{  x^2 + xy + y - sxy } 
   \right]~, \\
s F_{8}^{(9)} &=& \frac{8}{27}\pi^{2}   - \frac{88}{27}  
 - \frac{16}{9} \frac { s\ln(s) }{s - 1}  \no \\
&& +\frac{8}{9} \int_{0}^{1} dx \int_{0}^{1 - x}dy~\left[
   - 2 (2 - 3x - 3y) \ln\left( 1- s\frac{xy}{x^{2} + xy + y} \right) \right.
  \no \\ 
&& \quad \left.  +\frac{  2x^{3} + 4x^{2}y  + 2xy^{2} - x^{2} - xy + y
  -s( 2xy^{2} + 2x^{2}y  - 3xy + y)  }{ x^{2} + xy + y  - sxy } \right]~.\quad 
\eea

If we expand our results for small $s$, we recover  the results given 
in \cite{Asa1}.

\section{Non-perturbative contributions}
\label{nonperturbative}

\subsection{Generalities}

Non-perturbative contributions in $B\to X_s \ell^+\ell^-$
transitions can be divided into two main categories:
\begin{itemize}
\item $\Lambda_{\rm QCD}/m_b$ corrections in 
the relation between the partonic $b \to s \ell^+\ell^-$ amplitude 
and inclusive hadronic distributions;
\item non-perturbative 
effects associated with  the $c\bar{c}$ intermediate state:
$ B \to X_s c\bar{c} \to  X^\prime_s \ell^+\ell^-$.
\end{itemize}
The heavy-quark expansion, which led us to evaluate the 
first type of contributions, is rapidly convergent and leads 
to small corrections for {\em sufficiently inclusive} observables. 
A consistent treatment of the second type of effects 
requires to impose  {\em kinematical cuts} to avoid the large 
non-perturbative background of the narrow $c\bar{c}$ resonances.
These two requirements are somehow in conflict. 
As a result, we can perform reliable predictions
of $B\to X_s \ell^+\ell^-$ transitions,
both in the low- and in the high-$q^2$ regions,
but magnitude and error of the non-perturbative corrections
are enhanced with respect to their natural size. 

The enhancement of $\Lambda_{\rm QCD}/m_b$ corrections
is particularly sizeable in the high-$q^2$ region, 
because of two main drawbacks:
\begin{itemize}
\item the  $1/m_b$ expansion breaks down in the limit $q^2 \to m^2_b$ \cite{Alineu,BI};
\item the $q^2$ cut introduces a sizeable 
 effective correction linear in $1/m_b$, through the relation 
 between hadronic and partonic phase spaces. 
\end{itemize}

The first problem implies that in the high-$q^2$
region the differential distribution in $q^2$ 
cannot be predicted in perturbation theory.
This non-perturbative distribution 
has nothing to do with the so-called
shape function, or the kinetic energy distribution 
of the heavy quark inside the hadron \cite{BI}. However, similarly
to the latter, the  $q^2$ distribution near the end point
is a non-perturbative function, which must be determined from data.  
What can still be predicted with reasonable accuracy, 
for a sufficiently low cut-off $q_{\rm min}^2$, 
is the $q^2 > q_{\rm min}^2$ integral
(or the full inclusive distribution for $q^2 > q_{\rm min}^2$).

The second drawback is common to all observables 
that require kinematical cuts (in practice to any 
experimentally accessible inclusive observable in $B$ decays).
Since any kinematical cut on the final state must be expressed 
in terms of the hadron mass $M_{B}$, we cannot avoid the linear 
$1/m_b$ corrections that arises from  the relation 
\be
 M_{B} = m_b \left[1  + \frac{\bar \Lambda}{m_b}  - \frac{ \lambda_1 + 3 \lambda_2}{2 m_b^2} 
  + O\left(\frac{\Lambda_{\rm QCD}}{m_b^3},\alpha_s\right) \right]~.
\label{eq:lin}
\ee
This problem is substantially enhanced in the high-$q^2$ 
region because of the smallness of the available phase space: here the 
relative correction between the hadronic phase space 
[$\sim (M_{B}-\sqrt{q^2_{\rm min}})$] and the partonic 
one [$\sim (m_{b}-\sqrt{q^2_{\rm min}})$] becomes 
an $O(1)$ effect.
 
As we shall discuss in detail in the following, 
these two drawbacks fit within a common picture:
the heavy-mass expansion in the high-$q^2$ region 
is an effective expansion in inverse powers of 
\be
m^{\rm eff}_{\rm heavy}=m_b \times (1-\sqrt{s_{\rm min}})~,
\label{eq:m_eff}
\ee
rather than $m_b$. This expansion is justified,
but it converges less rapidly than the usual 
series in $\Lambda_{\rm QCD}/m_b$.

\subsection{$\Lambda^2_{\rm QCD}/m_b^2$ and $\Lambda^3_{\rm QCD}/m_b^3$ corrections}
\label{sect:mb23}

The $\Lambda_{\rm QCD}/m_b$ corrections can be systematically 
investigated in the framework of the heavy-quark expansion
and, in particular, by means of the  heavy-quark effective theory
(HQET) \cite{HQET}. The two main distributions, $R(s)$ and 
$A_{\rm FB}(s)$, are not affected by linear corrections
and the leading effects of $O(\Lambda^2_{\rm QCD}/m^2_b)$ can be 
described in terms of the two expectation values 
\be
\lambda_1=\frac{\langle B|\bar h(iD)^2h|B\rangle}{2 M_B}~,\qquad
\lambda_2=\frac{1}{6}\frac{\langle B|\bar hg\sigma\cdot Gh|B\rangle}{2 M_B}
=\frac{M^2_{B^*}-M^2_B}{4}~,
\ee
where $h$ is the heavy-quark field in the effective theory. The explicit expression 
of these corrections, which have been computed in Refs.~\cite{Alineu,BI}
(see also Ref.~\cite{Falk}), is
\bea
\delta_{1/m^2_b}R(s) &=& \frac{3\lambda_2}{2m^2_b}\Biggl(
\frac{\alpha_{\rm em}^2}{4\pi^2}\left|\frac{V_{ts}}{V_{cb}}\right|^2
\frac{1}{f(z)\kappa(z)}\Biggl\{  -(6+3s-5s^3)\frac{4|C^{\rm new}_7(s)|^2}{s} \no \\
&&\qquad  + (1-15s^2+10s^3)\left[ |C^{\rm new}_9(s)|^2+ |C^{\rm new}_{10}(s)|^2 \right] \no \\
&&\qquad  -4 (5+6s-7s^2) \Re\left[ C^{\rm new}_7(s) C^{\rm new}_9(s)^*\right]
 \Biggr\}+\frac{{g_{\lambda}(z)}}{f(z)}R(s)\Biggr)~, \qquad 
\label{eq:R_mb2} \\
\delta_{1/m^2_b}A_{\rm FB}(s) &=& \frac{3\lambda_2}{2m^2_b}\Biggl(
\frac{\alpha_{\rm em}^2}{4\pi^2}\left|\frac{V_{ts}}{V_{cb}}\right|^2
\frac{1}{f(z)\kappa(z)}\Biggl\{  s \mbox{Re} \left[ C^{\rm new}_{10}(s)^*  C^{\rm new}_9(s)  \right]
  (9+14s-15s^2)   \no \\
&& \qquad  + 2  \mbox{Re} \left[ C^{\rm new}_{10}(s)^* C^{\rm new}_7(s) \right] (7+10s-9s^2) \Biggr\}
   +\frac{{g_{\lambda}(z)}}{f(z)}A_{FB}(s)\Biggr)  \no \\  
&&    \qquad + \frac{4\lambda_1}{3m^2_b}\frac{s}{(1-s)^2} A_{\rm FB}(s)~.  \qquad
\label{eq:AFB_mb2}
\eea
In both cases we have taken into account also the $1/m^2_b$
terms arising from  the semileptonic normalization, namely
\bea
\Gamma(B\to X_c e\nu) &=& \frac{G^2_F m^5_b}{192\pi^3}|V_{cb}|^2
f(z)\kappa(z)\left[1+\frac{\lambda_1}{2m^2_b}-\frac{3\lambda_2}{2m^2_b}
  \frac{{ g_{\lambda}(z)}}{f(z)}\right]~, \\
{g_{\lambda}(z)}&=& { 3-8z+24z^2-24z^3+5z^4+12z^2\ln z~.}
\eea
This normalization is responsible for the absence of explicit $O(\Lambda_{\rm QCD}/m_b)$ 
corrections, since it cancels any explicit dependence from $m_b$, and 
it is also responsible for the absence of any dependence from  
the kinetic energy of the $b$-quark ($\sim\lambda_1$) in $R(s)$.

\begin{figure}[t]
\begin{center}
\epsfig{file=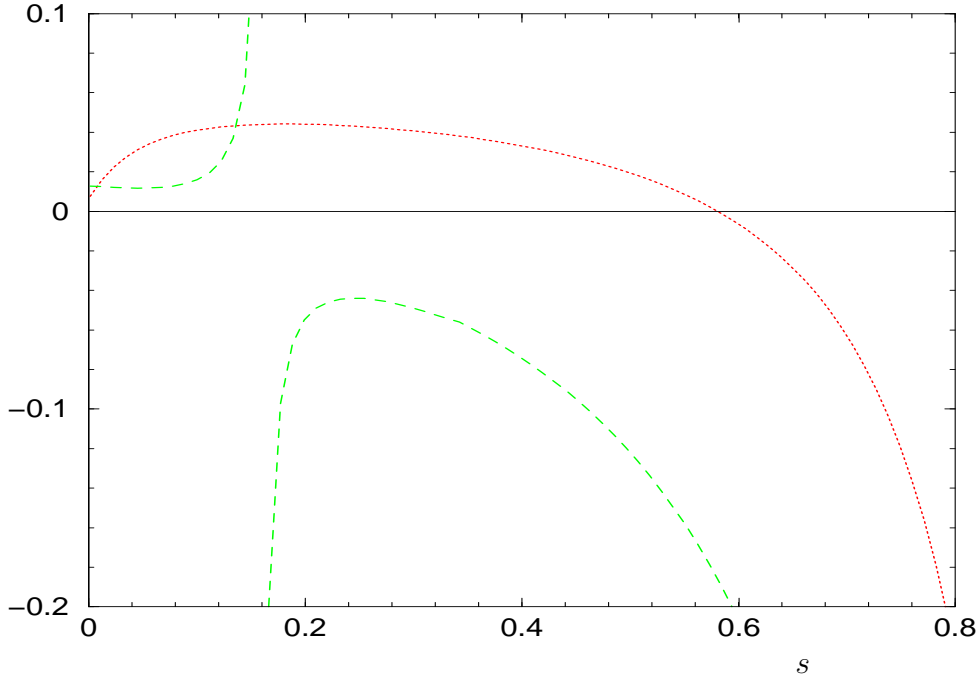,width=13cm,height=8.5cm}
\end{center}
\vskip -0.5  cm 
\hskip 13.0 cm $s$
\vskip 0.5 cm  
\caption{\footnotesize Relative corrections due $O(\Lambda^2_{\rm QCD}/m_b^2)$ effects: 
$\delta_{1/m^2_b}R(s)/R(s)$ (dotted) and $\delta_{1/m^2_b} A_{\rm FB}(s)/A_{\rm FB}(s)$ (dashed). }
\label{fig:plot_mbc}
\end{figure}

As can be seen in Fig.~\ref{fig:plot_mbc}, the relative corrections 
$\delta_{1/m^2_b}R(s)/R(s)$ and $\delta_{1/m^2_b} A_{\rm FB}(s)/A_{\rm FB}(s)$
are of the order of a few percent in the small $s$-region, apart from the 
obvious divergence in $\delta_{1/m^2_b} A_{\rm FB}(s)/A_{\rm FB}(s)$
due to the zero of $A_{\rm FB}(s)$ (the shift in the position of 
the zero amounts only to an increase of about $0.9\%$). 
However, in both cases the non-perturbative corrections 
become quite sizeable in the large $s$-region  
and the $1/m_b$ expansion breaks down close to the $s\to 1$ endpoint. 
The nature of this singularity has been discussed 
in detail in Ref.~\cite{BI}. As usual, the HQET cannot 
be applied in corners of the phase space 
of $O(\Lambda_{\rm QCD}/m_b)$, where the kinematics 
forces the final hadronic state to assume soft configurations. 
However, this particular case is rather different 
from the well-known examples of the 
photon-energy endpoint in  $B \to X_s \gamma$, or the 
lepton-energy endpoint in $B \to X_c \ell \nu$. There 
only the hadronic invariant mass is constrained to be 
semi-soft $(k^2 \sim m_b \Lambda_{\rm QCD})$
and the breakdown of the HQET is cured by means of a
resummation of singular terms which leads to the shape function, 
or the universal non-perturbative distribution of the 
$b$-quark kinetic energy  inside the hadron. 
Contrary to these examples, the kinematical constraint
corresponding to the dilepton invariant-mass end-point,
namely $m_b - \sqrt{q^2}=O(\Lambda_{\rm QCD})$, forces the hadronic
system to have {\em both} soft momentum $(k\sim \Lambda_{\rm QCD})$
and soft $(k^2 \sim \Lambda^2_{\rm QCD})$ invariant mass. 
This implies that no resummation can be applied 
and that the singularity has nothing to do with 
the kinetic-energy distribution of the $b$-quark \cite{BI}.

\medskip 

The fact that the $O(\Lambda_{\rm QCD}/m_b)$
corrections are not under control for $s\to 1$,
does not prevent us from performing  
reliable predictions for the partially integrated 
branching ratio (and FB asymmetry) in the high-$s$ 
region, provided we choose a {\em sufficiently low} cut-off $s_{\rm min}$.
The main issue is which is the maximal 
allowed value for $s_{\rm min}$.

Once we impose a constraint of the type $s> s_{\rm min}$, 
the inclusive sum on the hadronic
final state is limited to systems with 
invariant mass up to the effective 
scale $m^{\rm eff}_{\rm heavy}$ in (\ref{eq:m_eff}).
For this reason, in this partially integrated observables 
we expect an effective expansion ruled by 
inverse powers of $m^{\rm eff}_{\rm heavy}$, rather than $m_b$.
This naive expectation is confirmed by the detailed analysis 
of Ref.~\cite{Neubert}, applied $B \to X_u \ell\nu$ decays.
There the dilepton-invariant-mass cut necessary 
to avoid the $B \to X_c \ell\nu$ background 
leads to an effective expansion in inverse powers of 
$m_c$, rather than $m_b$ \cite{Neubert}.

\renewcommand\arraystretch{1.2}
\begin{table}[t]
\centering
\begin{tabular}{|c|c|c|c|}
\hline
   $\bar \Lambda$    &  $\lambda_1^{\rm eff}$(GeV$^2$) 
&  $\lambda^{\rm eff}_2$(GeV$^2$)      &  $\rho_1$(GeV$^3$)            
   \\ \hline
$0.40\pm 0.10$   &   $-0.15 \pm 0.10$    &  $0.12\pm 0.02$   &   $ 0.06 \pm 0.06$   
\\ \hline 
\end{tabular}
\caption{\footnotesize Input values of the HQET parameters used to estimate 
linear, quadratic and cubic corrections in the $1/m_b$ expansion.}
\label{tab:HQET}
\end{table}
\renewcommand\arraystretch{1}

Because of these general arguments, we expect 
 that the expansion should still be reliable, although with a  
slower convergence, for $s_{\rm min} \gsim 0.6$ 
(corresponding to $m^{\rm eff}_{\rm heavy}\gsim 1$~GeV). 
To address this issue in a more quantitative way, 
we shall look at the explicit expression of the 
$\Lambda_{\rm QCD}^3/m^3_b$ corrections in $R(s)$~\cite{Bauer}.
At this order we need to introduce seven new hadronic matrix elements.
Five of them lead only to a redefinition of the 
$\Lambda_{\rm QCD}^2/m^2_b$ couplings $\lambda_{1,2}$. 
In particular, the contributions proportional to 
${\cal T}_{1\ldots 4}$ and  $\rho_2$, in the notation of 
Ref.~\cite{Gremm}, are obtained from Eqs.~(\ref{eq:R_mb2}) 
and (\ref{eq:AFB_mb2}),  with the replacement~\cite{Bauer,Gremm}: 
\be
\lambda_{1} \to \lambda_{1}^{\rm eff} =  \lambda_{1} + \frac{ {\cal T}_1 +3 {\cal T}_2}{m_b}, \qquad
\lambda_{2} \to \lambda_{2}^{\rm eff} =  \lambda_{2} + \frac{ {\cal T}_3 +3 {\cal T}_4}{m_b} 
  -\frac{\rho_2}{m_b}~.
\ee
These terms do not spoil the convergence of the
$1/m_b$ expansions, independently of the $s_{\rm min}$ cut, 
provided  the naive chiral-counting 
expectation ${\cal T}_{1\ldots 4}\sim \rho_{1,2} \sim \Lambda^3_{\rm QCD}$
is respected. More delicate is the issue of the contributions
proportional to $\rho_1$ and $f_1$ \cite{Bauer}:
\bea
\delta_{1/m^3_b}R(s) &=& - \frac{\rho_1}{m^3_b}\Biggl( \frac{{ g_{\rho}(z)}}{6 f(z)}R(s) +
\frac{\alpha_{\rm em}^2}{4\pi^2}\left|\frac{V_{ts}}{V_{cb}}\right|^2
\frac{1}{f(z)\kappa(z)}\Biggl\{   \no \\
&& \left[\frac{5s^4+19s^3+9s^2-7s+22}{6(1-s)}+ 8 \Delta(f_1) \delta(1-s)\right]
   \frac{4|C^{\rm new}_7(s)|^2}{s}   \no \\
&& + \left[ \frac{ 10s^4+23s^3-9s^2+13s+11}{6(1-s)} + 8\Delta(f_1) \delta(1-s)\right]
   \left[ |C^{\rm new}_9(s)|^2+ |C^{\rm new}_{10}(s)|^2 \right] \no \\
&& +   4\left[ \frac{-3s^3+17s^2-s+3}{2(1-s)}+ 8\Delta(f_1) \delta(1-s)\right]
   \Re\left[ C^{\rm new}_7(s) C^{\rm new}_9(s)^*\right]
 \Biggr\} \Biggr)~, \qquad 
\label{eq:R_mb3} 
\eea
where 
\beq
g_{\rho}(z) = 77-88z+24z^2-8z^3+5z^4+48\ln z +36 z^2\ln z
\eeq
arises from the semileptonic normalization and $\Delta(f_1)$ 
is a local contribution that  cures the singularity 
of $\int_{0}^s ds' R(s')$ for 
$s\to 1$.\footnote{~The cut-off-dependent coupling $\Delta(f_1)$ 
is related to the cut-off-independent parameters $f_1$ and $\rho_1$, 
defined as in Ref~\cite{Bauer}, by  the relation
$\int_{0}^1 ds [ 1/(1-s) + \Delta(f_1) \delta(1-s) ] = - f_1/\rho_1 $. }
When integrated in the high-$s$ region,  Eq.~(\ref{eq:R_mb3}) leads 
to a huge coefficient for the  $\rho_1/m^3_b$ correction, 
much larger than the already sizeable $\lambda_2/m^2_b$ term~\cite{Bauer}. 
However, everything looks very reasonable,  once we introduce the 
effective scale in (\ref{eq:m_eff}). For $s_{\rm min} \approx 0.6$ we find 
\be
\int_{s_{\rm min}}^{1} ds~R(s) =  \left[ 1 - \frac{ 1.6 \lambda_2}{m_b^2(1-\sqrt{s_{\rm min}})^2}
+ \frac{1.8 \rho_1 + 1.7 f_1}{m_b^3(1-\sqrt{s_{\rm   min}})^3} \right] \times 
  \int_{s_{\rm min}}^{1} ds~\left. R(s)\right|_{m_b\to \infty}~,
\label{eq:mb3f}
\ee
which perfectly confirms our expectation of an effective expansion 
in inverse powers of $m^{\rm eff}_{\rm heavy}$.
According to the input values in Table~\ref{tab:HQET}, which 
are consistent with recent experimental determinations  
(see e.g.~Ref.~\cite{CKMBook}), and setting $f_1=0$,\footnote{~To 
fix this quantity requires more restricting information, but 
we assume its contribution is within the present uncertainty.}
the numerical size of the term between square brackets 
in (\ref{eq:mb3f}) is  $[1-0.08 \pm 0.08]$.

\subsection{The $\bar \Lambda/m_b$  correction}
As anticipated, even though  $R(s)$ and $A_{\rm FB}(s)$ are not 
explicitly affected by linear corrections in the $1/m_b$ 
expansion,  the physical observables defined in terms of a $q^2$ cut 
are sensitive to the $\bar \Lambda/m_b$ term via the relation (\ref{eq:lin}).
This term -- or equivalently the uncertainty on the value of $m_b$ --
represent at present the largest source of non-perturbative 
uncertainty in the high-$q^2$ region. 
Choosing as reference cut the value $s_{\rm min} = 0.6$, 
the physical observable defined in terms of $q^2_{\rm min}$ 
can be written as 
\bea
R_{\rm cut}(q^2_{\rm min})
 &=& \int_{ q^2 > q^2_{\rm min} } dq^2 
\frac{ d \Gamma(B\to X_s \ell^+\ell^-)}{\Gamma(B\to X_c e \nu)} \no \\
 &=&  \left\{ 1- 6.2 \left(\frac{q^2_{\rm min}}{m_b^2}-0.6 \right) + 
O \left[ \left(\frac{q^2_{\rm min}}{m_b^2}-0.6 \right)^2 \right] \right\}\times \int_{0.6}^{1} ds~R(s)~,
\eea
which implies 
\be
\frac{\delta R_{\rm cut}}{ R_{\rm cut}} \approx 7.4 \frac{\delta m_b }{m_b}~.
\ee
This means that an error $\delta m_b=0.1$~GeV (corresponding to 
the uncertainty on $\bar \Lambda$ in Table~\ref{tab:HQET}), 
leads to a $\approx 15\%$  error on $R_{\rm cut}$. 

\subsection{$1/m_c^2$ non-perturbative corrections}
\label{sect:cc}

The second class of non-perturbative effects relevant in 
$B\to X_s \ell^+ \ell^-$ decays are the long-distance 
corrections related to $c\bar c$ intermediate states.
These originate from the non-per\-tur\-ba\-tive interactions
of the $c\bar c$ pair in the process $B\to X_s c\bar c \to X_s
\ell^+ \ell^-$. If the dilepton invariant mass is near the first two  
$J^{PC}=1^{--}$ $c\bar c$ resonances ($\Psi$ and $\Psi'$), 
this effect is very large  and shows up as a peak in $R(s)$. 
However, one can easily eliminate this background
by suitable kinematical cuts. More delicate is the estimate of 
the long-distance effects away from the resonance peaks.

An interesting approach which avoid double-counting 
problems is the one proposed in Ref.~\cite{KS} (KS approach).
Here, in order to take into account charm rescattering,
the correction to $C_9$ induced by $b\to c\bar{c}s$ operators is 
estimated by means of experimental data on 
 $\sigma(e^+e^-\to c\bar c$ hadrons)  using a dispersion relation.
To be more specific, the function $h(z,s)$ appearing in (\ref{hzs})  
is replaced by 
\begin{equation}\label{hKS}
h(z,s) \longrightarrow  h(z,0) + {s\over 3} P \int_{s_c}^{\infty}
ds' \frac{ R^{c\bar{c}}_{\rm had} (s') }{s'(s'-s)}
+ i\frac{\pi}{3} R^{c\bar{c}}_{\rm had} (s)~,
\end{equation}
where $R^{c\bar{c}}_{\rm had} (s) = 
\sigma(e^+e^-\to c\bar c)/\sigma(e^+e^-\to \mu^+\mu^-)$ and
$s_c$ is the $c\bar{c}$ threshold.
This method is exact only in the limit 
where the $\bar B\to X_s c\bar c$ transition can be 
factorized into the product of $\bar{s}b$
and $\bar{c}c$ colour-singlet currents 
(i.e. {\it non-factorizable} effects are not included). 
The non-perturbative corrections estimated using this approach are extremely 
small in the perturbative windows $s<0.25$ and $s>0.6$
\cite{Gudrun}. For the integrated branching ratios one finds 
an increase of $1-2\%$ in the low-$s$ region, while the 
effect in the high-$s$ region is far below
the uncertainty of the $1/m_b$ corrections. 

A systematic and model-independent way to estimate the non-factorizable
$c\bar c$ long-distance effects far from the resonance region
is obtained by means of an expansion in inverse powers of the 
charm-quark mass \cite{Rey,Chen}.
This approach, originally proposed in \cite{VOL} to evaluate 
similar effects in $B\to X_s\gamma$ decays, has the advantage 
of dealing only with partonic degrees of freedom. In this framework 
the leading non-perturbative corrections to $R(s)$ and 
$A_{\rm FB}$ turn out to be $O(\Lambda^2_{\rm QCD}/m_c^2)$;
their explicit expressions can be found in the appendix. 
Since  the 
factorizable corrections vanish for $s\to 0$, 
the $O(\Lambda^2_{\rm QCD}/m_c^2)$ effect
is expected to be the dominant long-distance 
contribution for small values of the dilepton invariant mass. 
In this region the relative magnitude
is very small (at the $1$  or $2 \%$ level) and opposite in 
sign to the factorizable KS correction. The $O(\Lambda^2_{\rm QCD}/m_c^2)$
calculation should be reliable also above the resonance 
region ($s >0.6$), where the effect is again very small.
Similar comments apply to the long-distance corrections for $A_{\rm FB}(s)$.

\section{Phenomenological analysis}
\subsection{Branching ratio and dilepton invariant-mass spectrum}
\begin{figure}[p]
\vskip  -0.5  cm
\begin{center}
\hskip  1.5 cm
\epsfig{file=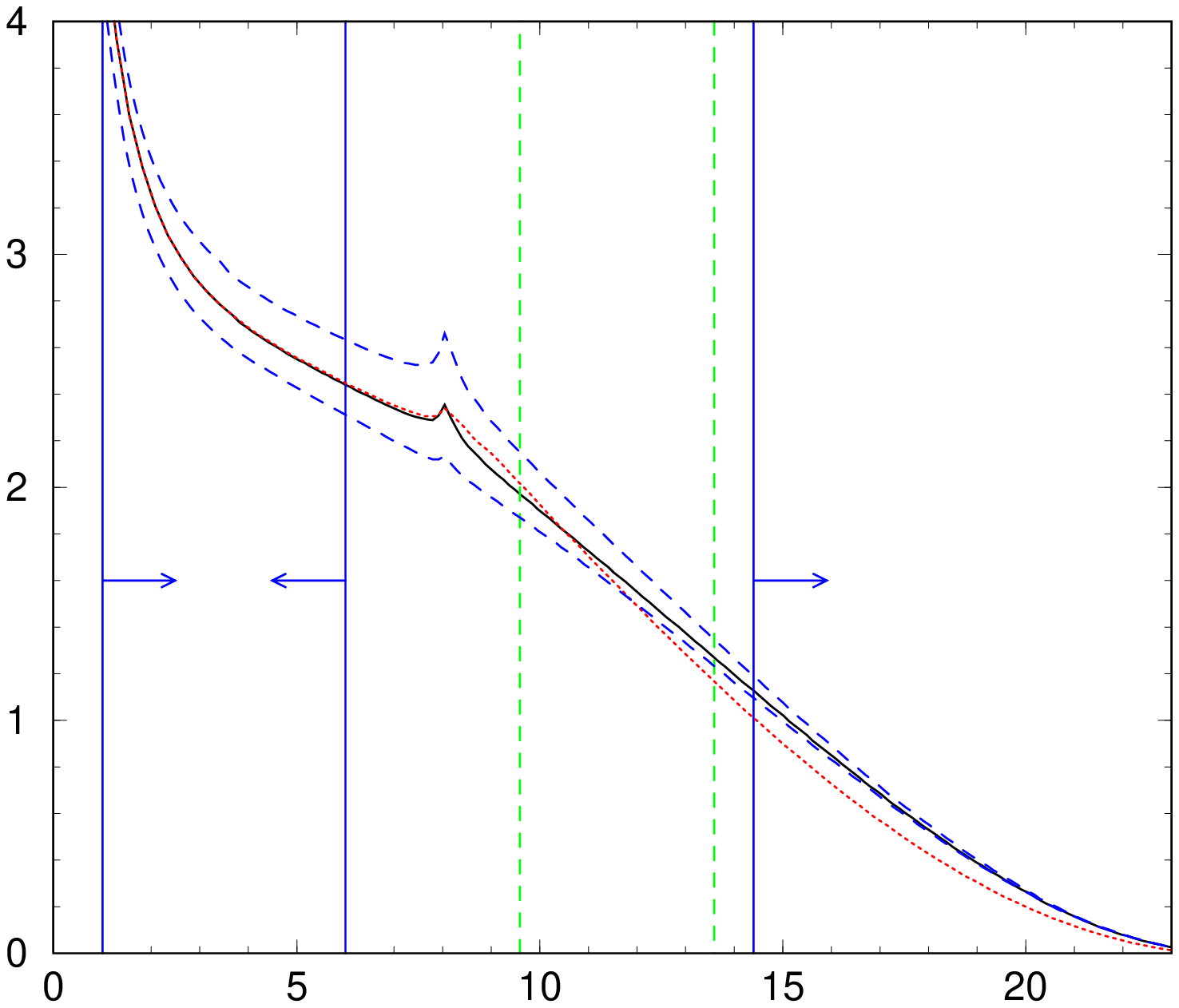,width=13cm,height=8.5cm}
\end{center}
\vskip -8.7 cm  $\quad 10^7 \times\displaystyle\frac{d{\cal B}}{dq^2}$ 
\vskip  0.2 cm  $\quad~({\rm GeV}^{-2})$
\vskip  7.0  cm 
\begin{center}
\hskip  1.5 cm
\epsfig{file=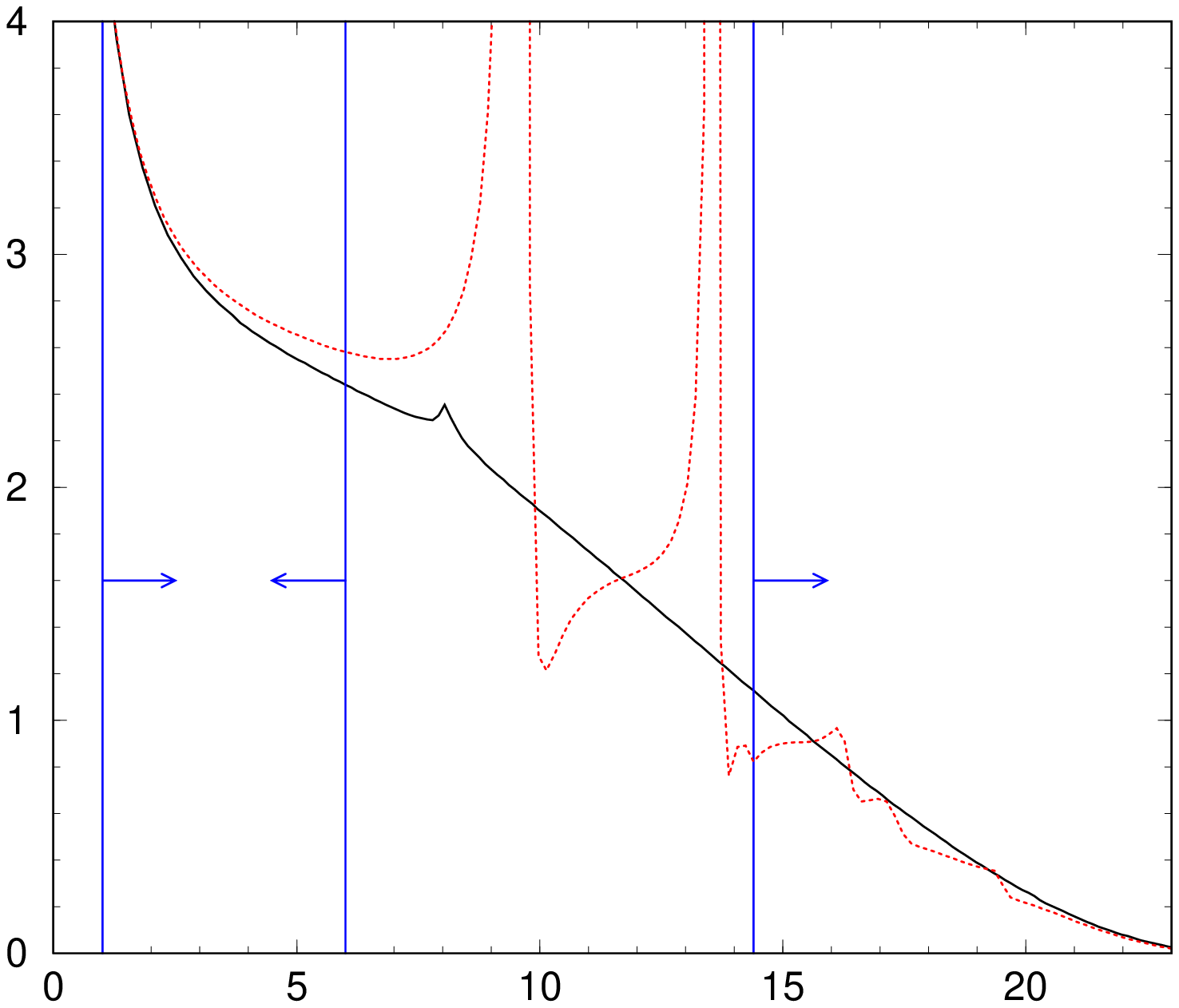,width=13cm,height=8.5cm}
\end{center}
\vskip -8.7 cm  $\quad 10^7 \times\displaystyle\frac{d{\cal B}}{dq^2}$ 
\vskip  0.2 cm  $\quad~({\rm GeV}^{-2})$
\vskip  6.7  cm 
\hskip  11.8 cm $q^2~({\rm GeV}^2)$
\vskip  0.2 cm  
\caption{\footnotesize NNLL predictions of $d{\cal B}(B\to X_s \ell^+\ell^-)/dq^2$.
Upper plot: pure partonic result with full $m_c$ dependence 
computed in this work for $\mu$=5~GeV (full line), $\mu$=2.5 and 10~GeV
(dashed lines); $q^2/(4m^2_c)$-expanded result by Asatrian et al. \cite{Asa1}, for $\mu$=5 GeV,
extrapolated to the full $q^2$ range (dotted line); the dashed vertical lines 
indicate the positions of the first two narrow $\Psi$ resonances. 
Lower plot: partonic result with full $m_c$ dependence for $\mu$=5~GeV with 
(dotted line) or without (full line) factorizable $c\bar c$ corrections 
computed in the KS approach (see Sect.~\ref{sect:cc}). 
All other inputs are fixed to the central values in Table~\ref{tab:main_inputs}}
\label{fig:plot_R}
\end{figure}

The final results of this work concerning the dilepton invariant-mass 
spectrum is summarized in Fig.~\ref{fig:plot_R}. In the upper plot 
we compare our 
un-expanded result, without any non-perturbative correction, to 
the expanded result of Ref.~\cite{Asa1}. As can be noted, the 
expanded result provides a perfect approximation to the full 
calculation up to about the $c\bar c$ threshold. 
This is of course a good cross-check 
of Ref.~\cite{Asa1} and an important test of our method, which turns 
out to be essential to provide a reliable prediction in the high-$q^2$ region.
The Fig.~\ref{fig:plot_R} also shows that the expanded result
differs significantly from the exact NNLL result above the threshold 
as expected.
Note that the scale dependence in the  high-$q^2$ region is 
very small, therefore our NNLL result provides an excellent 
level of accuracy for the pure partonic calculation.
The lower plot in Fig.~\ref{fig:plot_R} provides an
illustration of the non-perturbative effects induced by 
$c\bar c$ intermediate states (evaluated using the KS approach, see Sect.~\ref{sect:cc}).
As can be noted, the point-by-point corrections in $q^2$
are quite sizeable in the high-$q^2$ window. As discussed 
in the previous section, only the $q^2$-integral can be 
predicted reliably in this region.

\renewcommand\arraystretch{1.3}
\begin{table}[t]
\centering
\begin{tabular}{|c|c|c|}
\hline
   $m_b=(4.9\pm 0.1)$~GeV    
&  $m_c/m_b=0.29 \pm 0.02$
&  $\mu=\left(5.0^{+5.0}_{-2.5}\right)$~GeV  
 \\ \hline
\end{tabular}
\vskip 0.3 cm 
\begin{tabular}{|c|c|c|c|}
\hline
   $\alpha_s(M_Z)= {0.119}$ 
&  $\alpha_{\rm em}=1/128$ 
&  $|V^*_{tb} V_{ts}/V_{cb}|=0.97$
&  $\overline{m_t} (m_t) = 167$~GeV
 \\ \hline
\end{tabular}
\caption{\footnotesize Main input values used in the numerical analysis.}
\label{tab:main_inputs}
\end{table}
\renewcommand\arraystretch{1}

Before discussing the numerical predictions for the integrated 
branching ratios, we wish to emphasize that low- and high-$q^2$ regions 
have complementary virtues and disadvantages. Taking into account the 
discussion in the previous section, we can summarize the main points 
as follows:
\begin{itemize}
\item
{\em Virtues of the low-$q^2$ region:} reliable $q^2$ spectrum; 
small $1/m_b$ corrections; sensitivity to the interference of $C_7$
and $C_9$; high rate.
\item
{\em Disadvantages of the low-$q^2$ region:} difficult to perform 
a fully inclusive measurement (severe cuts on the dilepton 
energy and/or the hadronic invariant mass); long-distance effects 
due to processes of the type $B \to \Psi X_s \to  X_s + X^\prime \ell^+\ell^-$ 
not fully under control; non-negligible scale 
and $m_c$ dependence.
\item
{\em Virtues of the high-$q^2$ region:} negligible scale 
and $m_c$ dependence due to the strong sensitivity to $|C_{10}|^2$;
easier to perform a fully inclusive measurement (small 
hadronic invariant mass); negligible long-distance effects of the type 
$B \to \Psi X_s \to  X_s + X^\prime \ell^+\ell^-$.
\item
{\em Disadvantages of the high-$q^2$ region:} $q^2$ spectrum  
not reliable; sizeable $1/m_b$ corrections; low rate.
\end{itemize}
Given this situation, we believe that future experiments should 
try to measure the branching ratios in both regions and 
report separately the two results. These two measurements 
are indeed affected by different systematic uncertainties 
(of theoretical nature) and provide different 
short-distance information.

In order to obtain theoretical predictions which can be confronted with
experiments, it is necessary to define the two regions with appropriate 
cuts in $q^2$ (and not in the partonic variable $s$, as done in most 
of the previous literature). 
Concerning the low-$q^2$ window, 
we propose as reference interval the range
$q^2 \in [1,6]~{\rm GeV}^2$.
The lower bound on $q^2$ is not essential, but it proposed in order to cut 
a region where there is no new information with respect to $B\to X_s \gamma$ 
and where we cannot trivially combine electron and muon modes.
The higher cut is essential to decrease the uncertainty associated to the 
$c\bar c$ threshold.

Taking into account the input values in Table~\ref{tab:main_inputs}, 
the NNLL prediction within the SM for this low-$q^2$ window is:
\bea
R^{\rm low}_{\rm cut} &=&
 \int_{ 1~{\rm GeV}^2 }^{6~{\rm GeV}^2 } dq^2 
\frac{ d \Gamma(B\to X_s \ell^+\ell^-)}{\Gamma(B\to X_c e \nu)}
= 1.48 \times 10^{-5} \nonumber \\
 && \times   
\Biggl[ 1  \pm 8\% \big|_{\Gamma_{\rm sl}}  \pm 6.5\% \big|_{\mu}   \pm 2\% \big|_{m_c}    
\pm 3\% \big|_{m_b({\rm cuts})} 
  + (4.5 \pm 2)\% \big|_{1/m^2_b}  -(1.5 \pm 3)\% \big|_{c\bar c}  \Biggl] \nonumber \\
 &=&   (1.52 \pm 0.18 )\times 10^{-5}~. 
\label{eq:lowR} 
\eea
Between square brackets we have reported all the uncertainties 
and non-perturbative corrections discussed in this work, evaluated according to the input values 
in Table~\ref{tab:HQET} and~\ref{tab:main_inputs}. 
The error denoted by $\Gamma_{\rm sl}$ corresponds to the theoretical 
uncertainty implied by the $\Gamma(B\to X_c e \nu)$ normalization
which, in turn, is dominated by the uncertainty on $m_c$. 
In principle, alternative normalizations 
such as the one proposed in Ref.~\cite{MisiakG} could be used to 
reduce this error. In any case, this uncertainty should be regarded 
as a parametric error which can be improved 
by additional independent measurements. The small error denoted by 
$m_c$ correspond to the $m_c$-dependence of $\Gamma(B\to X_s \ell^+\ell^-)$
only (ignoring the normalization): as can be noted, this is almost negligible.

\begin{figure}[t]
\vskip  -0.5  cm
\begin{center}
\hskip  1.5 cm
\epsfig{file=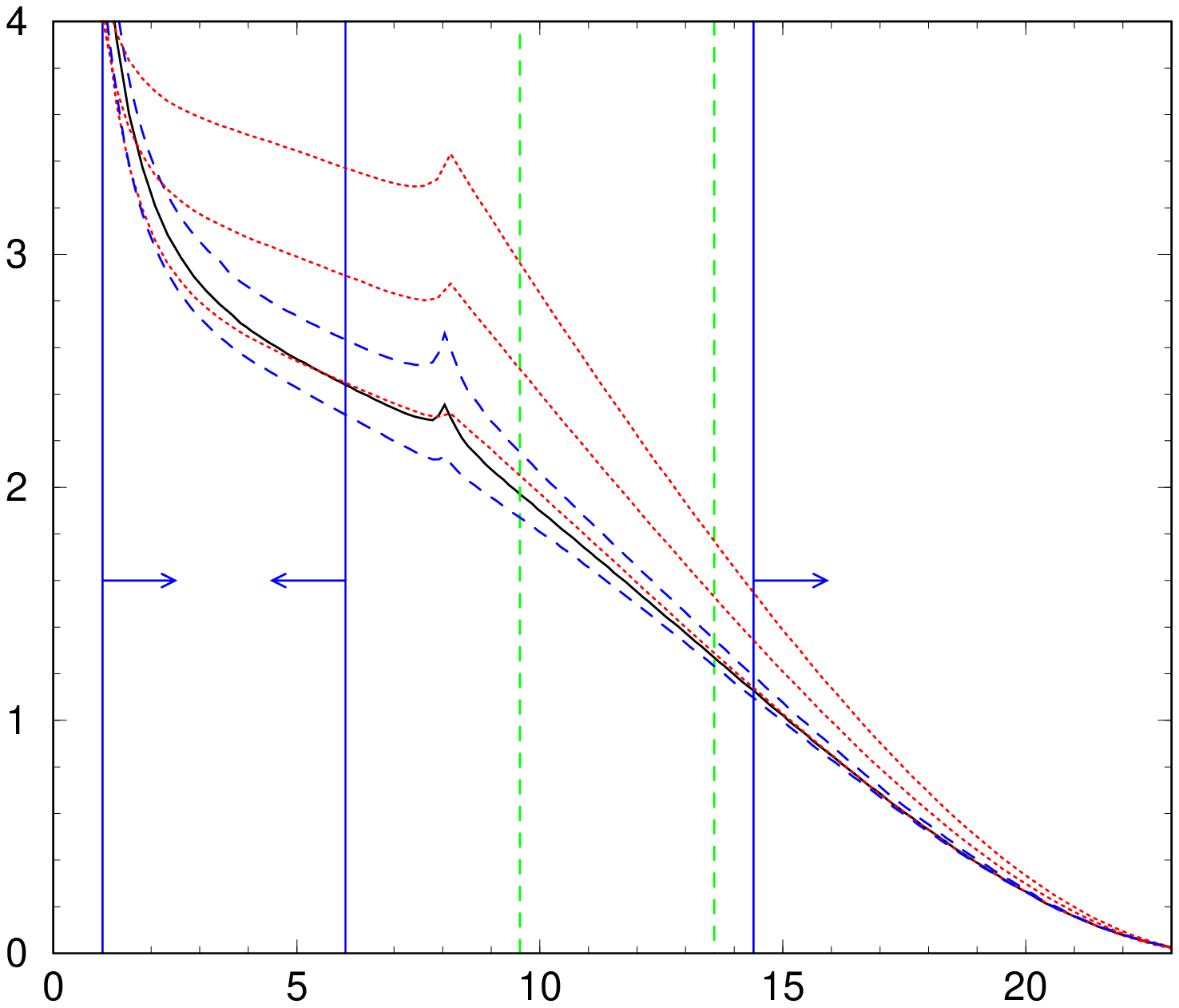,width=13cm,height=8.5cm}
\end{center}
\vskip -8.7 cm  $\quad 10^7 \times\displaystyle\frac{d{\cal B}}{dq^2}$ 
\vskip  0.2 cm  $\quad~({\rm GeV}^{-2})$
\vskip  6.7  cm 
\hskip  11.8 cm $q^2~({\rm GeV}^2)$
\vskip  0.2 cm  
\caption{\footnotesize 
Complete NNLL result for $d{\cal B}(B\to X_s \ell^+\ell^-)/dq^2$ for $\mu$=5~GeV (full line), 
$\mu$=2.5 and 10~GeV (dashed lines), vs. the partial NNLL result obtained 
by neglecting the $F^{(7,9)}_{1,2,8}$ functions computed in this work, 
 for $\mu=2.5$~GeV (lower dotted line), $\mu=5$~GeV (middle dotted line), and 
$\mu=10$~GeV (upper dotted line). }
\label{fig:plot_Rmu}
\end{figure}

As already pointed out, our calculation of the matrix elements provides 
a completely independent check of the results of Ref.~\cite{Asa1} 
for the low-$q^2$ region. In particular, we confirm the reduction 
of the scale uncertainty from $\pm 13\%$ to $\pm 6.5\%$,
in the branching ratio, once these NNLL corrections are included 
(see Fig.~\ref{fig:plot_Rmu}). The difference in the central value 
of $R^{\rm low}_{\rm cut}$, compared to the result of Ref.~\cite{Asa1},
is entirely due a different definition of this observable 
and to differences in the input values.

\medskip

Concerning the high-$q^2$ window, we propose as reference cut $q^2 > 14.4 {\rm GeV}^2$,
which leads to the following first NNLL prediction of the high dilepton 
mass spectrum:
\bea
R^{\rm high}_{\rm cut} &=&
 \int_{ q^2> 14.4~{\rm GeV}^2 } dq^2 
\frac{ d \Gamma(B\to X_s \ell^+\ell^-)}{\Gamma(B\to X_c e \nu)} = 
 4.09 \times 10^{-6} \nonumber \\
&& \times 
   \Biggl[ 1  \pm 8\% \big|_{\Gamma_{\rm sl}}  \pm 3\% \big|_{\mu}  
  + 0.15 \left(\frac{m_b - 4.9~{\rm GeV}}{0.1~{\rm GeV}}\right)
- (8 \pm 8)\% \big|_{1/m^{(2,3)}_b}  \pm 3 \% \big|_{c\bar c}  \Biggl] \nonumber \\
&=&   (3.76 \pm 0.72 )\times 10^{-6}~.\quad 
\label{eq:highR}
\eea
Here the explicitly indicated $m_b$ dependence induces the largest uncertainty. 
At present this is about $15\%$. However, significant improvements can be expected 
in the near future in view of more precise data on other inclusive semileptonic
distributions. Note that, as anticipated, in this region the pure perturbative 
uncertainties due to scale and $m_c$ dependence are very small (the latter has not
been explicitly indicated being below 1\%). The impact of the NNLL corrections 
computed in this work for the high $q^2$ region 
is a $13\%$ reduction of the central value      and a significant 
reduction of the perturbative scale dependence (from $\pm 13\%$ to
$\pm 3\%$, see Fig.~\ref{fig:plot_Rmu}). 

As also shown in  Fig.~\ref{fig:plot_Rmu}, this reduction brings the 
central value of the full NNLL prediction very close to the 
partial NNLL result obtained without the $F^{(7,9)}_{1,2,8}$ functions, 
provided in the latter case the renormalization scale 
is set around $2.5$ GeV.      This observation, which has already 
been made in Ref.~\cite{Ali_extrap} for the low-$q^2$ window, 
applies remarkably well also in the high $q^2$ region. 
Obviously, our full NNLL calculation provides a fundamental 
ingredient to justify this procedure for the central value 
and, especially, to obtain a clear estimate of the residual
scale uncertainty.

\medskip

It must be stressed that two non-negligible source of uncertainties have not 
been explicitly included in Eqs.~(\ref{eq:lowR}) and  (\ref{eq:highR}): 
the error due to $m_t$ (and the high-energy
QCD matching scale) and the error due to $\alpha_{\rm em}$ (or better 
the error due to  higher-order electroweak and electromagnetic effects). The first type of
uncertainty has been discussed in detail in \cite{MisiakBobeth}, and it 
amounts to $\approx 6 \%$. As far as the uncertainty on higher-order 
electroweak corrections is concerned, the error is also expected 
to be at the level of a few percent, but a consistent estimate of 
these effects 
is beyond the scope of this work.\footnote{~Choosing as reference value for 
$\alpha_{\rm em}$ the value at the electroweak scale, we should 
have minimized the impact of the dominant electroweak matching corrections. 
After this work has been completed, a complete analysis of
higher-order electroweak corrections has been presented 
in Ref.~\cite{Paolonew2}. } 
 
\medskip

Using the world average 
$\Gamma(B\to X_c e \nu)=(10.74\pm0.24)\%$ \cite{hfag}, we finally obtain:
\bea
{\cal B}(B\to X_s \ell^+\ell^-;~q^2 \in [1,6]~{\rm GeV}^2 ) &=& (1.63 \pm 0.20) \times 10^{-6}~, 
 \label{eq:pred1} \\
{\cal B}(B\to X_s \ell^+\ell^-;~q^2 > 14.4~{\rm GeV}^2 )    &=& (4.04 \pm 0.78) \times 10^{-7}~. 
 \label{eq:pred2}
\eea
At the moment, these two predictions cannot be directly compared with experimental data.
To facilitate the comparison with the recent results by BELLE and BABAR \cite{Bellebsll,Babarbsll},
we present also an estimate 
for the {\em interpolated partonic rate}.   The latter 
is defined as the integral of the partonic rate 
(full line in Fig.~\ref{fig:plot_R})
over the full $q^2$ spectrum (including the resonance region), 
starting from  $q^2_{\rm min}=4 m^2_\mu$:
\beq
{\cal B}(B\to X_s \ell^+\ell^-; ~q^2 > 4 m^2_\mu )  = (4.6 \pm 0.8) \times 10^{-6}~. 
\label{eq:BR_full}
\eeq
The central value of this estimate, based on our NNLL calculation,  
takes into account the power corrections 
in the clean perturbative windows, while its error includes a guestimate  of the  systematic uncertainty 
due to the extrapolation (evaluated using the KS approach). 
The difference of our central value with respect to a  similar estimate
for the interpolated partonic rate presented in Ref.~\cite{Ali_extrap} 
is almost entirely due to parametric differences in the input values 
(most notably, to the choice of $\alpha_{\rm em}$) and is well within the
theoretical errors given in (\ref{eq:BR_full}).
 
Our estimate in (\ref{eq:BR_full}) compares well with the  
recent experimental world average \cite{Nakao}:
\beq
{\cal B}( B \to X_s \ell^+\ell^- )^{\rm exp} = (6.2 \pm 1.1 {}^{+1.6}_{-1.3} ) \times 10^{-6}~.
\label{eq:bsll_exp}
\eeq
In view of data with higher statistics, we stress once more the importance 
of a future comparisons with the more clean and more interesting predictions in 
Eqs.~(\ref{eq:pred1}) and (\ref{eq:pred1}). Contrary to Eq.~(\ref{eq:BR_full}), 
the theoretical errors in both (\ref{eq:pred1}) and (\ref{eq:pred1})
could be systematically improved in the near future.

\subsection{Forward-backward asymmetry}

\begin{figure}[t]
\begin{center}
\hskip  1.5 cm
\epsfig{file=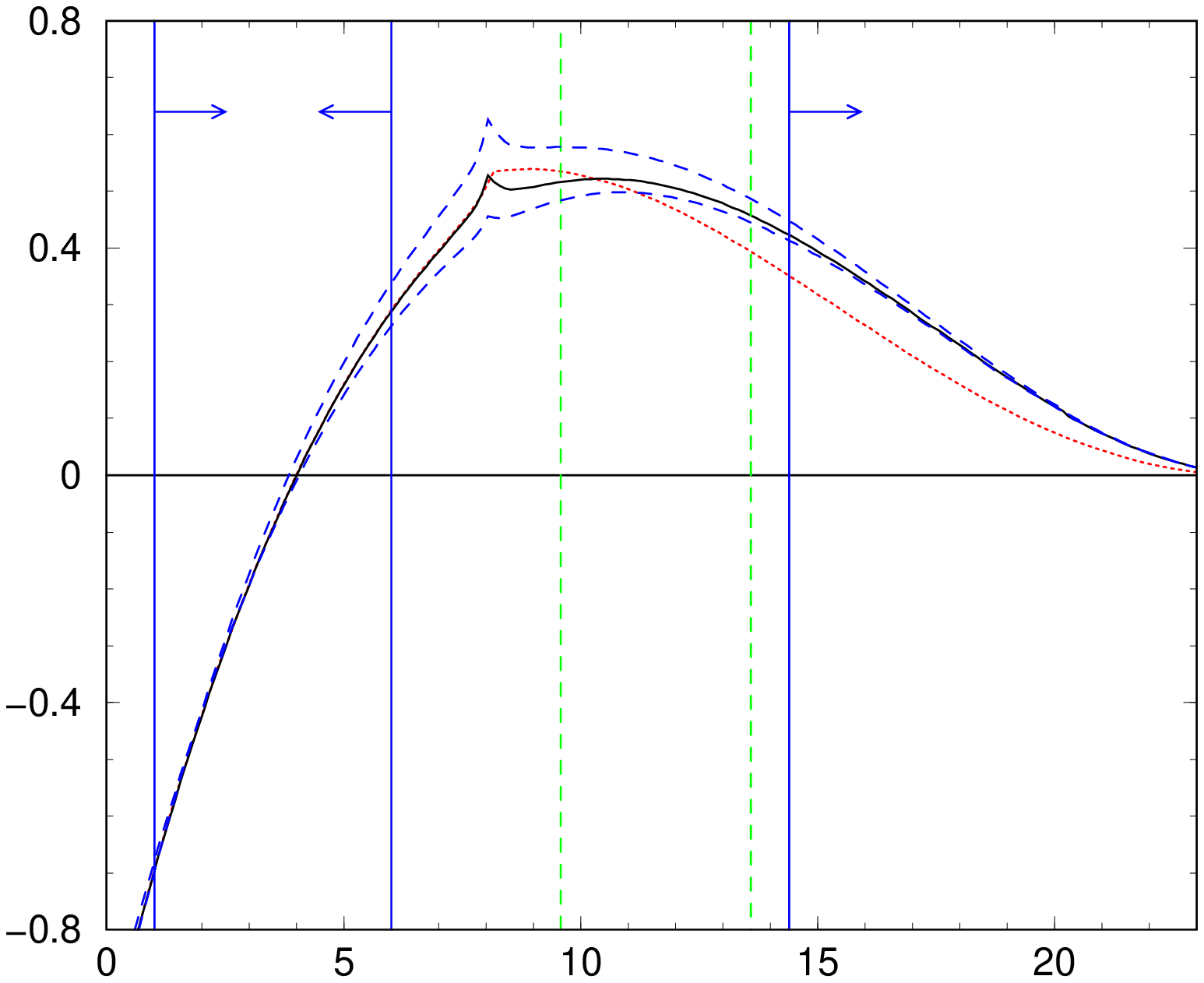,width=13cm,height=8.5cm}
\end{center}
\vskip -8.7 cm  $\quad 10^7 \times\displaystyle\frac{d{\cal A}_{\rm FB}}{dq^2}$ 
\vskip  0.2 cm  $\quad\ \ ({\rm GeV}^{-2})$
\vskip  7.0  cm 
\hskip 12.0 cm $q^2~({\rm GeV}^2)$
\vskip 0.5 cm  
\caption{\footnotesize NNLL perturbative contributions to the un-normalized FB asymmetry; 
notations as in the upper plot of Fig.~\ref{fig:plot_R}.}
\label{fig:AFB}
\end{figure}

\begin{figure}[t]
\begin{center}
\epsfig{file=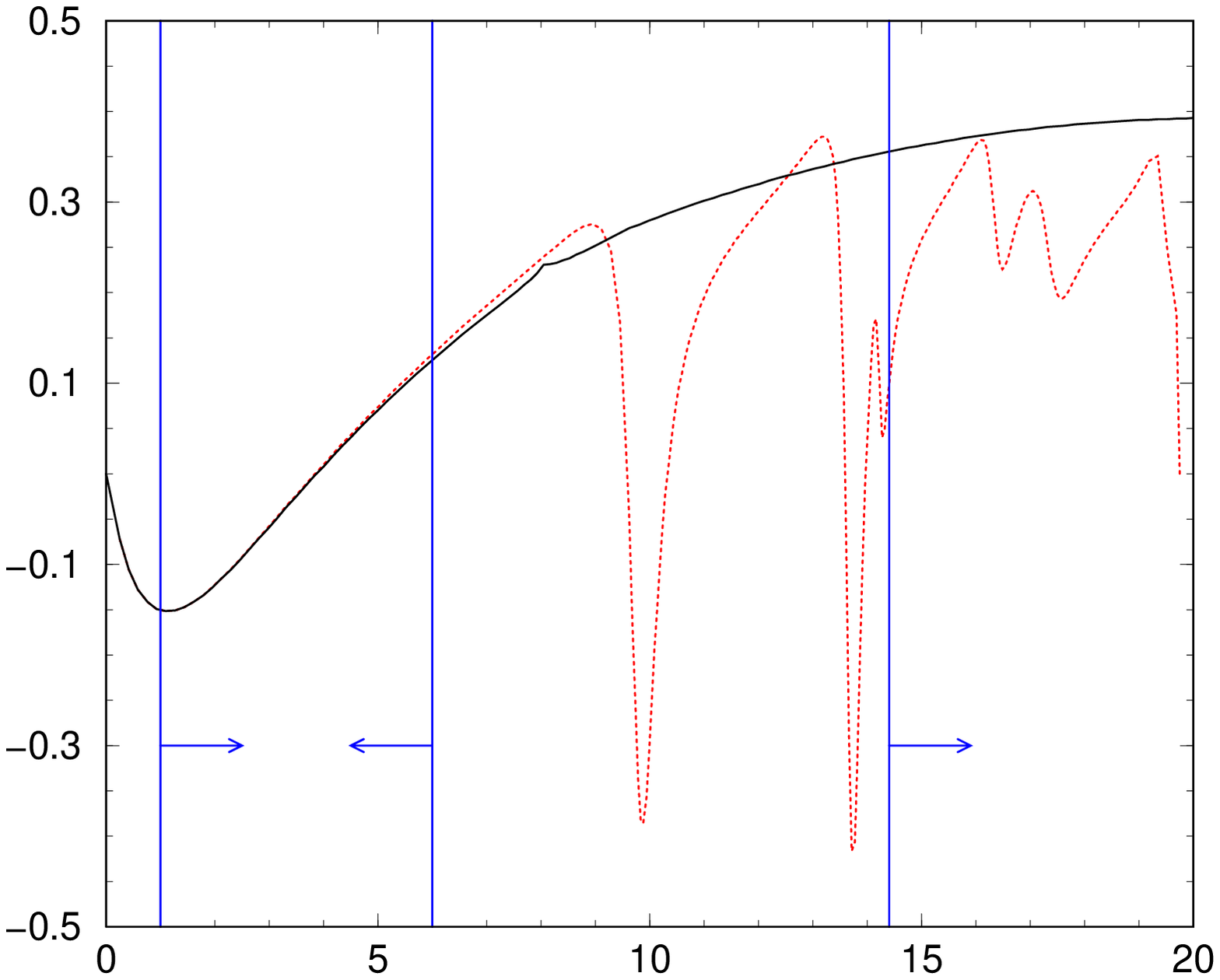,width=13cm,height=8.5cm}
\end{center}
\vskip -8.5 cm  $\displaystyle \bar{\cal A}_{\rm FB}(q^2)$ 
\vskip  7.8  cm 
\hskip  12.3 cm $q^2~({\rm GeV}^2)$
\vskip 0.5 cm  
\caption{\footnotesize NNLL perturbative contributions to the normalized FB asymmetry; notations as in 
the lower plot of Fig.~\ref{fig:plot_R}.}
\label{fig:AFB2}
\end{figure}

\noindent 
The summary plots for the lepton forward-backward asymmetry are shown 
in Fig.~\ref{fig:AFB} and \ref{fig:AFB2}. In  Fig.~\ref{fig:AFB} we plot 
the un-normalized differential asymmetry, defined by 
\bea
\frac{ d {\cal A}_{\rm FB}(q^2)}{ d q^2}  &=&  \int_{-1}^1 d\cos\theta_\ell ~
 \frac{d^2 {\cal B} ( B\to X_s \ell^+\ell^-)}{d q^2  ~ d\cos\theta_\ell}
\mbox{sgn}(\cos\theta_\ell) \no \\
&=&  \frac{ {\cal B} (B\to X_ce\bar{\nu}) }{ m_b^2 } A_{\rm FB} \left( \frac{q^2}{m_b^2} \right)~,
\eea
while in  Fig.~\ref{fig:AFB2} we plot the (adimensional) normalized differential asymmetry,
defined by 
\be
\bar {\cal A}_{\rm FB}(q^2) = \frac{1}{d {\cal B} ( B\to X_s \ell^+\ell^-) /d q^2  }
  \int_{-1}^1 d\cos\theta_\ell ~
 \frac{d^2 {\cal B} ( B\to X_s \ell^+\ell^-)}{d q^2  ~ d\cos\theta_\ell}
\mbox{sgn}(\cos\theta_\ell)~.
\ee
Most of the comments concerning the errors and the 
complementary of low- and high-$q^2$ windows discussed 
before holds also for the forward-backward asymmetry. 

In the low-$q^2$ region the  most interesting observable 
is not the integral of the asymmetry, which is very small
due to the change of sign, but the position of the zero. 
As discussed by several authors (see e.g. Ref.~\cite{Adrian1,Asa3}),
this is one of the most precise predictions 
(and one of the most interesting SM tests) 
in rare $B$ decays. 
Denoting by $q^2_0$ the position of the zero, 
and showing explicitly only the uncertainties 
and non-perturbative effects larger than 0.5\%,
we find at the NNLL order 
\beq
q^2_0 = 0.161 \times m_b^2 \times \left[ 1 + 0.9\% \big|_{1/m_b^2}  
\pm 5\% \big|_{\rm NNNLL} \right] = (3.90 \pm 0.25)~{\rm GeV^2}~.
\label{eq:q0}
\eeq
As already pointed out in Ref.~\cite{Adrian1}, in this case the $\mu$ 
dependence is accidentally small and does not provides a conservative 
estimate of higher-order QCD corrections. The $5\%$ error in (\ref{eq:q0})
has been estimated comparing the result within the ordinary LL counting 
and within the modified perturbative ordering proposed in Ref.~\cite{Asa1} 
(see Sect.~\ref{sect:ordering}). The central value, as well as all the 
other central values in this work, is obtained using the 
modified ordering of Ref.~\cite{Asa1}.

\medskip

In the high-$q^2$ window the FB asymmetry does not change sign,
therefore its integral represents an interesting observable.
In order to minimize non-perturbative and normalization
uncertainties it is more convenient to consider a 
normalized integrated asymmetry. Applying the same 
$q^2$ cut as in (\ref{eq:highR}), we define
\beq
(\bar{\cal A}_{\rm FB})_{\rm cuts}^{\rm high} =  \left[
 \int_{ q^2> 14.4~{\rm GeV}^2 } dq^2~ \frac{ d {\cal B}_{\rm FB}(q^2)}{ d q^2} \right]^{-1}
 \int_{ q^2> 14.4~{\rm GeV}^2 } dq^2~ \frac{ d {\cal A}_{\rm FB}(q^2)}{ d q^2}~.
\eeq
All parametric and perturbative uncertainties are very 
small in this observable at the NNLL order level. 
On the other hand, despite a partial 
cancellation, this ratio is still affected by a considerable amount 
of $\Lambda^2_{\rm QCD}/m_b^2$ and $\Lambda^3_{\rm QCD}/m_b^3$ 
corrections (which represent the by-far dominant 
source of uncertainty). Taking into account the expressions
in Sect.~\ref{sect:mb23} and separating the contributions 
of the various operators, we find 
\bea
(\bar{\cal A}_{\rm FB})_{\rm cuts}^{\rm high} &=& 0.42 \times \left[ 1 
 -(0.17 \pm 0.11)_{\lambda_1} - (0.42 \pm 0.07)_{\lambda_2} -(0.08\pm 0.08)_{\rho_1} \right] \no \\
&=& 0.14 \pm 0.06~.
\eea

\section*{Acknowledgements} 
We thank M.~Walker for providing us with partial results of the calculation 
in \cite{Asa1} which made possible a detailed comparison with our results.
We also thank P.~Gambino for useful discussions. 
This work is partially 
supported by the EC-Contract HPRN-CT-2002-00311 (EURIDICE)
and the U.S. Department of Energy (DOE).

\setcounter{figure}{0}
\setcounter{equation}{0}
\renewcommand{\theequation}{A.\arabic{equation}}
\renewcommand{\thefigure}{A\arabic{figure}}

\section*{Appendix 1: Auxiliary functions
}

\begin{itemize}
\item  The function $h(z)$ describing next-to-leading order QCD corrections
to the semileptonic decay [see Eq.~(\ref{semi})] is given by \cite{NirNir}:
\bea
\label{nirfunction}
    h(z) = & -(1-z^2) \, \left( \frac{25}{4} - \frac{239}{3} \, z + \frac{25}{4} \, z^2 \right) + z \, \ln(z)
        \left( 20 + 90 \, z -\frac{4}{3} \, z^2 + \frac{17}{3} \, z^3 \right)
        \nonumber \\ &
        + z^2 \, \ln^2(z) \, (36+z^2) + (1-z^2) \, \left(\frac{17}{3} - \frac{64}{3} \, z +
        \frac{17}{3} \, z^2 \right) \, \ln (1-z)
        \nonumber \\ &
        - 4 \, (1+30 \, z^2 + z^4) \, \ln(z) \ln(1-z) -(1+16 \, z^2 +z^4)  \left( 6 \, \mbox{Li}(z) - \pi^2 \right)
        \nonumber \\ &
        - 32 \, z^{3/2} (1+z) \left[\pi^2 - 4 \, \mbox{Li}(\sqrt{z})+ 4 \, \mbox{Li}(-\sqrt{z})
        - 2 \ln(z) \, \ln \left( \frac{1-\sqrt{z}}{1+\sqrt{z}} \right) \right]~.
\eea

\item In the counterterms to the two-loop matrix elements we use the following 
functions:

\bea
  B(a)  &=& \int_0^1 \ln(1 - x (1-x) a)  \nonumber\\
 B_x(a)  &=& \int_0^1 x \ln( 1 - x (1-x) a )  \nonumber\\
B_{xx}(a)  &=& \int_0^1 x^2 \ln( 1 - x (1-x) a) \nonumber\\   
 B_2(a) &=& \int_0^1  (\ln( 1 - x (1-x) a ))^2  \nonumber\\
 B_{2x}(a) &=& \int_0^1  x (\ln( 1 - x (1-x) a ))^2  \nonumber\\
B_{2xx}(a) &=& \int_0^1   x^2 (\ln( 1 - x (1-x) a ))^2  \nonumber\\
\eea
For the explicit expressions one should check $a  = a + I \epsilon$  
in order to remain on the correct Riemann sheet. Further functions are needed
for the unexpanded counterterms:
\bea
  C(a)  &=& \int_0^1 2 (x (1-x) a)  /(1 - x (1 - x) a ) \nonumber\\
 C_x(a)  &=& \int_0^1 2 x (x (1-x) a)  /(1 - x (1 - x) a ) \nonumber\\
C_{xx}(a)  &=& \int_0^1 2 x^2 (x (1-x) a)  /(1 - x (1 - x) a ) \nonumber\\
C_2(a) &=&  \int_0^1  4   \ln( 1 - x (1-x) a ) (x (1-x) a)  /(1 - x (1 - x) a) \nonumber\\
C_{2x}(a) &=&  \int_0^1  4  x  \ln( 1 - x (1-x) a ) (x (1-x) a)  /(1 - x (1 - x) a) \nonumber\\
C_{2xx}(a) &=&  \int_0^1  4 x^2  \ln( 1 - x (1-x) a ) (x (1-x) a)  /(1 - x (1 - x) a) \nonumber\\
\eea

\end{itemize}

\section*{Appendix 2: Effective Wilson coefficients} 

\begin{figure}[t]
\begin{center}
\epsfig{file=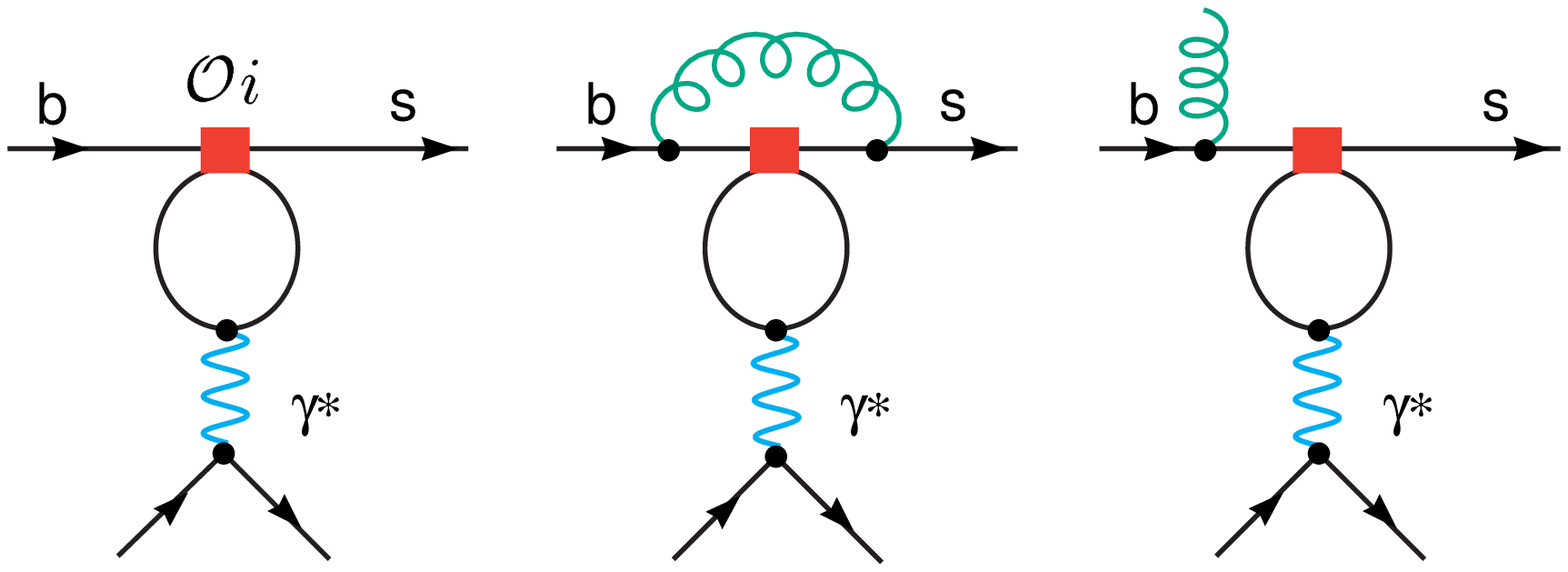,width=10cm}
\end{center}
\caption{\footnotesize Examples of virtual and bremsstrahlung 
contributions of the four-quark operators ${\cal O}_{1\ldots 6}$ that 
are taken into account by the redefinition
of the Wilson coefficients in Eq.~(\ref{effWilson}).
\label{bsll01}}
\end{figure}

\begin{figure}[t]
\begin{center}
\epsfig{file=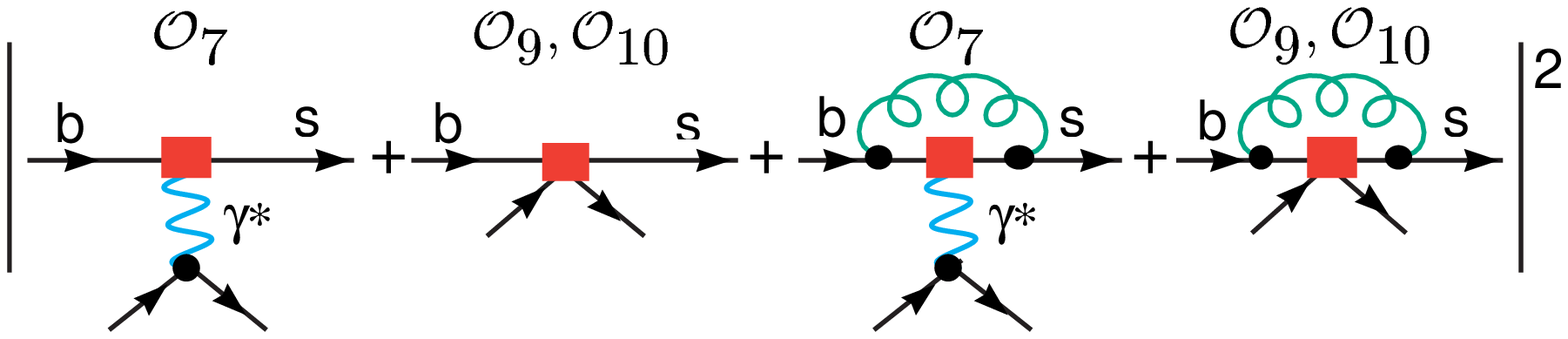,width=13.3cm}
\epsfig{file=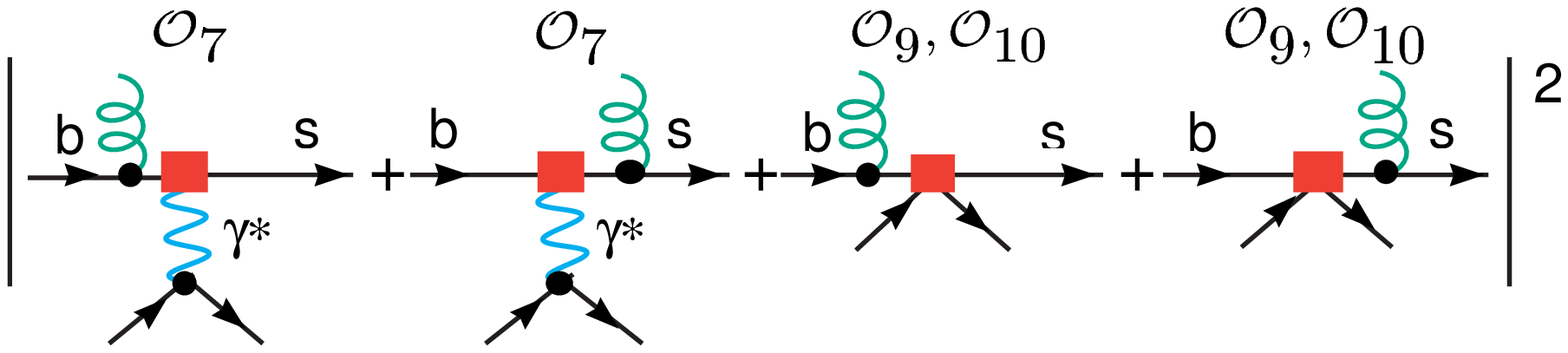,width=13.3cm}
\end{center}
\caption{\footnotesize Virtual (up) and real (down) QCD corrections generating the 
terms $\tau_i$ and $\sigma_i$.
\label{bsll03}}
\end{figure}

The effective coefficients $\widetilde{C}_{7-10}^{\rm eff}$ 
appearing in Eq.~(\ref{effmod}) are defined in our notation as,
\bea
\widetilde{C}_7^{\rm eff} &=& \frac{4 \pi}{\alpha_s(\mu)} C_7(\mu)
-\frac{1}{3} C_3(\mu) -\frac{4}{9} C_4(\mu)
-\frac{20}{3} C_5(\mu) -\frac{80}{9} C_6(\mu)\nonumber \\
\widetilde{C}_8^{\rm eff} &=& \frac{4 \pi}{\alpha_s(\mu)} C_8(\mu)
 + C_3(\mu) -\frac{1}{6} C_4(\mu)
+ 20  C_5(\mu) -\frac{10}{3} C_6(\mu) \nonumber\\
\widetilde{C}_9^{\rm eff}(s) &=& 
  \frac{4 \pi}{\alpha_s(\mu)} C_9(\mu)
+ \sum_{i=1}^6 C_i(\mu) \gamma^{(0)}_{i9}  \ln\left(\frac{m_b}{\mu}\right)
\nonumber\\ &+& 
\frac{4}{3} C_3(\mu)+ \frac{64}{9} C_5(\mu)+ \frac{64}{27} C_6(\mu)
\nonumber\\
&+& h\left(z,s \right) \left( 
\frac{4}{3} C_1(\mu) + C_2(\mu)   
+ 6 C_3(\mu) + 60 C_5(\mu) \right)
\nonumber\\ &+& h(1,s) \left(  
-\frac{7}{2} C_3(\mu)-\frac{2}{3} C_4(\mu)-38 C_5(\mu)-\frac{32}{3} C_6(\mu) \right)
\nonumber\\ &+& h(0,s) \left(  
-\frac{1}{2} C_3(\mu)-\frac{2}{3} C_4(\mu)- 8 C_5(\mu)-\frac{32}{3} C_6(\mu) \right)
\nonumber\\ &\equiv& A_9 + h(z,s) T_9 + h(1,s) U_9 + h(0,s) W_9 \nonumber\\
\widetilde{C}_{10}^{\rm eff} &=& 
 \frac{4 \pi}{\alpha_s(\mu)}   C_{10}(\mu)~,
\label{effWilson}
\eea
where 
\bea
h(z,s) &=&  - \frac{4}{9} \ln(z) + \frac{8}{27} + \frac{16}{9}\frac{z}{s} 
              - \frac{2}{9} \left( 2+\frac{4\, z}{s} \right)
              \sqrt{\left|\frac{4\,z-s}{s}\right|} \times \no\\
       &\times& \left\{ \ba{ll}
 2 \arctan \sqrt{\frac{s}{4\,z-s}} & \mbox{for} \, s < 4\,z\, ,   \\ 
 \ln \left(\frac{\sqrt{s} + \sqrt{s - 4\,z}}{\sqrt{s} - 
\sqrt{s - 4\,z}} \right) -i\,\pi \qquad &\mbox{for}\,  s > 4\,z  \, . \ea \right.
\label{hzs}
\eea
Note that specific one- and two-loop and matrix-element contributions  
of the four-quark operators ${\cal O}_{1-6}$ 
(including the corresponding bremsstrahlung contributions) such as the one 
shown in Fig.~\ref{bsll01} are   included by 
the redefinition of the 
Wilson coefficients $C_7$, $C_9$ and $C_{10}$
given in (\ref{effWilson}). In fact, using this redefinition,    
the bremsstrahlung and virtual corrections that are 
shown in Fig.~\ref{bsll03} (see next subsection) 
automatically take  these effects into 
account. The Wilson coefficients $C_i$ in (\ref{effWilson}), 
which are needed to NNLL precision, 
are presented in \cite{MisiakBobeth,Asa1}. For completeness we quote
them here again:

\begin{table}[t]
\begin{center}
\begin{tabular}{| c | c | c | c |} \hline 
    &$\mu=2.5$ GeV& $\mu=5$ GeV  & $\mu=10$ GeV\\ \hline
    $\alpha_s$ & $0.267$&  $0.215$ & $  0.180$\\ \hline
    $C_1^{(0)}$ & $-0.697$ & $-0.487$ & $-0.326$\\ \hline
    $C_2^{(0)}$ & $1.046 $& $ 1.024$ &  $ 1.011$\\ \hline
    $\big(\widetilde{C}_7^{{\rm eff}\,(0)},~\widetilde{C}_7^{{\rm eff}\,(1)}\big)$ & $ (-0.360,~0.031)$ & $(-0.321,~0.019)$
& $ (-0.287,~0.008)$\\ \hline
    $\widetilde{C}_8^{{\rm eff}\,(0)}$ &$ -0.164 $& $ -0.148$& $ -0.134$\\ \hline
    $\big(A_9^{(0)},~A_9^{(1)}\big)$ & $ (4.241,~-0.170) $& $(4.129,~0.013)$ & $
(4.131,~0.155)$ \\ \hline
    $\big(T_9^{((0))},~T_9^{(1)}\big)$ & $ (0.115,~0.278) $& $ (0.374,~0.251)$ &
$ (0.576,~0.231)$ \\ \hline
    $\big(U_9^{(0)},~U_9^{(1)}\big)$ & $ (0.045,~0.023)$ & $ (0.032,~0.016)$ & $
(0.022,~0.011)$ \\ \hline
    $\big(W_{9}^{(0)},~W_{9}^{(1)}\big)$ & $(0.044,~0.016)$ & $ (0.032,~0.012)$
& $ (0.022,~0.009)$ \\ \hline
    $\big(\widetilde{C}_{10}^{{\rm eff}\,(0)},~\widetilde{C}_{10}^{{\rm eff}\,(1)}\big)$ & $ (-4.372,~0.135)$ &  $(-
4.372,~0.135)$ &  $ (-4.372,~0.135)$ \\ \hline
\end{tabular}
\caption{\footnotesize Numerical values of Wilson coefficients of (\ref{effWilson}) for
 three different values of $\mu$; the $\alpha_s$ expansion of the 
terms are defined by 
$C_i^{\rm eff} = C_i^{{\rm eff}\,(0)} + C_i^{{\rm eff}\,(1)} + ...$\,  .}
\end{center}
\end{table}
 
\section*{Appendix 3: IR virtual and bremsstrahlung corrections}

The universal $O(\alpha_s)$ bremsstrahlung and the corresponding
infrared (IR) virtual corrections which can be absorbed into
the Wilson coefficients (see (\ref{effmod})) were calculated
in \cite{Asa2,Adrian1,Asa3}  and given by
\bea
  \sigma_9(s) &=& \sigma(s) + \frac{3}{2}~, \qquad\qquad  
  \sigma_7(s) \ = \ \sigma(s) + \frac{1}{6} - \frac{8}{3}\ln\left( \frac{\mu}{m_b} \right)~, 
\no\\
  \sigma(s) &=& - \frac{4}{3} \mbox{Li}_2(s) - \frac{2}{3} \ln(s) \ln(1-s)
  -\frac{2}{9}\pi^2 -\ln(1-s)-\frac{2}{9}(1-s)\ln(1-s)~. \qquad 
\label{eq:sigmai}
\eea

The remaining (finite) non-universal bremsstrahlung are encoded
in rate (see \ref{NNLLDIMS})) and FB asymmetry (see \ref{NNLLAFB})).
We note that we have chosen the universal functions $\sigma_i$ in 
(\ref{eq:sigmai}) such that
the non-universal contributions to the rate, namely 
$\tau_{77},\tau_{99},\tau_{79}$, vanish in the limit $s \rightarrow 1$:
\bea
\tau_{77}(s) &=& - \frac{2}{9(2+s)} \left[   2(1-s)^2\ln(1-s)
+\frac{6s(2-2s-s^2)}{(1-s)^2}\ln(s) +\frac{11-7s-10s^2}{(1-s)} \right]~, 
\no\\
 \tau_{99}(s) &=& 
-\frac{4}{9(1+2s)} \left[ 2(1-s)^2\ln(1-s) 
+\frac{3s(1+s)(1-2s)}{(1-s)^2}\ln(s)+\frac{3(1-3s^2)}{1-s} \right]~, \no 
\\
\tau_{79}(s) &=&  - \frac{4(1-s)^2}{9s} \ln(1-s)
-\frac{4s(3-2s)}{9(1-s)^2}\ln(s) -\frac{2(5-3s)}{9(1-s)}~,
\label{eq:tau_R}  \\
&& \no \\
\tau_{710}(s) &=& -\frac{5}{2} + \frac{1}{3 (1-3s)} - \frac{1}{3} 
\frac{s (6-7 s) \ln(s)}{(1-s)^2}- \frac{1}{9} 
\frac{(3-7 s + 4 s^2) \ln(1-s)}{s} + \frac{f_7 (s)}{3} \no \\
\tau_{910}(s)  &=& -\frac{5}{2} + \frac{1}{3(1-s)} - \frac{1}{3} 
 \frac{s (6-7s) \ln(s)}{((1-s)^2)} - \frac{2}{9} \frac{(3-5 s + 2 s^2) 
\ln(1-s)}{s}  + \frac{f_ 9 (s)}{3} 
\eea
where 
\bea
f_7 (s) &=& \frac{1}{6 (s-1)^2 } 
\Big\{
  24(1+13s -4s^2) {\rm Li}_2(\sqrt{s})
+ 12(1-17s+6s^2) {\rm Li}_2(s) +6s(6-7s) \ln(s)
\no \\ && 
+24(1-s)^2\ln(s)\ln(1-s) + 12(-13+16s-3s^2)[\ln(1-\sqrt{s})-\ln(1-s)]
\no \\ &&  +39 -2\pi^2 +252s -26\pi^2s
+21s^2+8\pi^2s^2 -180\sqrt{s} -132s\sqrt{s} 
\Big\}~,  \\ && \no  \\ 
f_9 (s) &=&  -\frac{1}{6 (s-1)^2 } 
\Big\{
 48 s (-5+ 2s) {\rm Li}_2(\sqrt{s})  
+ 24(-1+7s-3s^2)  {\rm Li_2}(s) + 6s (-6+7s) \ln(s)
\no \\ && 
-24(1-s)^2 \ln(s)\ln(1-s) +24 (5-7s+2s^2) 
[\ln(1-\sqrt{s})-\ln(1-s)] 
\no \\ &&  -21-156s+20\pi^2s
+9s^2-8\pi^2s^2+120 \sqrt{s}+48s\sqrt{s}
\Big\}~.
\eea

\section*{Appendix 4: Complete set of scalar integrals}

In \cite{method} it was shown that there are ten 
linear combinations of the  integrals  
\begin{equation}
\tilde {\cal P}^{a \, b}_{\alpha_1 \, \alpha_2 \, \alpha_3} (m_1,m_2,m_3;k^2)
  =
    \int d^{n}p\,d^{n}q\, 
       \frac{(p \cdot k)^a (q \cdot k)^b}{
             [(p+k)^{2}+m_{1}^{2}]^{\alpha_{1}} \,
             (q^{2}+m_{2}^{2})^{\alpha_{2}} \,
             (r^{2}+m_{3}^{2})^{\alpha_{3}} \, , 
            }
\end{equation}
with $a+b \le3$, which are sufficient for treating {\it all} two-loop Feynman 
diagrams which one can encounter in renormalizable theories. 
Because the ultravilet behaviour of the functions ${\cal H}_i$ is logarithmic 
only, one finds simple {\it finite} integral representations: 
\begin{eqnarray}
   {\cal H}_1 
   & = &
    \pi^4 
    \left[
      \frac{1}{2 \epsilon^2}
    + \frac{1}{2 \epsilon} ( 1 - 2 \gamma_{m_1} )
          - \frac{1}{2}
          + \frac{\pi^2}{12}
          - \gamma_{m_1}
          + \gamma_{m_1}^2
          + h_1
   \right]
  \nonumber \\
   {\cal H}_2 
   & = &
    \pi^4 k^2
    \left[
    - \frac{1}{2 \epsilon^2}
    - \frac{1}{2 \epsilon} ( \frac{1}{2} - 2 \gamma_{m_1} )
          + \frac{13}{8}
          - \frac{\pi^2}{12}
          + \frac{\gamma_{m_1}}{2}
          - \gamma_{m_1}^2
          - h_2
   \right]
  \nonumber \\
   {\cal H}_3 
   & = &
    \pi^4 k^2
    \left[
      \frac{1}{4 \epsilon^2}
    + \frac{1}{2 \epsilon} ( \frac{1}{4} - \gamma_{m_1} )
          - \frac{13}{16}
          + \frac{\pi^2}{24}
          - \frac{\gamma_{m_1}}{4} 
          + \frac{\gamma_{m_1}^2}{2} 
          + h_3
   \right]
  \nonumber \\
   {\cal H}_4 
   & = &
    \pi^4 (k^2)^2
    \left[
      \frac{3}{8 \epsilon^2}
    - \frac{1}{2 \epsilon} \frac{3 \gamma_{m_1}}{2} 
          - \frac{175}{96}
          + \frac{\pi^2}{16}
          + \frac{3 \gamma_{m_1}^2}{4} 
          + \frac{3}{4} h_4
   \right]
  \nonumber \\
   {\cal H}_5 
   & = &
    \pi^4 (k^2)^2
    \left[
    - \frac{3}{16 \epsilon^2}
    + \frac{1}{2 \epsilon} \frac{3 \gamma_{m_1}}{4} 
          + \frac{175}{192}
          - \frac{\pi^2}{32}
          - \frac{3 \gamma_{m_1}^2}{8} 
          - \frac{3}{4} h_5
   \right]
  \nonumber \\
   {\cal H}_6 
   & = &
    \pi^4 (k^2)^2
    \left[
      \frac{1}{8 \epsilon^2}
    + \frac{1}{2 \epsilon} ( \frac{1}{24} - \frac{\gamma_{m_1}}{2}  )
          - \frac{19}{32}
          + \frac{\pi^2}{48}
          - \frac{\gamma_{m_1}}{24} 
          + \frac{\gamma_{m_1}^2}{4} 
          + \frac{3}{4} h_6
   \right]
  \nonumber \\
   {\cal H}_7
   & = &
    \pi^4 (k^2)^3
    \left[
    - \frac{1}{4 \epsilon^2}
    + \frac{1}{2 \epsilon} ( \frac{5}{24} + \gamma_{m_1} )
          + \frac{287}{192}
          - \frac{\pi^2}{24}
          - \frac{5 \gamma_{m_1}}{24} 
          - \frac{\gamma_{m_1}^2}{2} 
          - \frac{1}{2} h_7
   \right]
  \nonumber \\
   {\cal H}_8
   & = &
    \pi^4 (k^2)^3
    \left[
      \frac{1}{8 \epsilon^2}
    -  \frac{1}{2 \epsilon} ( \frac{5}{48} + \frac{\gamma_{m_1}}{2}  )
          - \frac{287}{384}
          + \frac{\pi^2}{48}
          + \frac{5 \gamma_{m_1}}{48} 
          + \frac{\gamma_{m_1}^2}{4} 
          + \frac{1}{2} h_8
   \right]
  \nonumber \\
   {\cal H}_9 
   & = &
    \pi^4 (k^2)^3
    \left[
    - \frac{1}{12 \epsilon^2}
    +\frac{1}{2 \epsilon} ( \frac{1}{24} + \frac{\gamma_{m_1}}{3}  )
          + \frac{95}{192}
          - \frac{\pi^2}{72}
          - \frac{\gamma_{m_1}}{24} 
          - \frac{\gamma_{m_1}^2}{6} 
          - \frac{1}{2} h_9
   \right]
  \nonumber \\
   {\cal H}_{10} 
   & = &
    \pi^4 (k^2)^3
    \left[
      \frac{1}{16 \epsilon^2}
    - \frac{1}{2 \epsilon} ( \frac{1}{96} + \frac{\gamma_{m_1}}{4}  )
          - \frac{283}{768}
          + \frac{\pi^2}{96}
          + \frac{\gamma_{m_1}}{96} 
          + \frac{\gamma_{m_1}^2}{8} 
          + \frac{1}{2} h_{10}
   \right]  
   \; \; .
\end{eqnarray}
$d=4 - 2 \epsilon$ is the space-time dimension, and 
$\gamma_m=\gamma + \log(\pi m^2/\mu_1^2)$. 
The special functions $h_i$ which appear in the formulae above are 
the finite part in the $1/\epsilon$ expansion of ${\cal H}_i$
and cannot be 
further integrated into well-studied functions, such as the 
familiar polylogarithms:
\begin{eqnarray}
    h_1(m_1,m_2,m_3;k^2) & = &  \int_0^1 dx \,
                                \tilde{g} (x)
  \nonumber \\
    h_2(m_1,m_2,m_3;k^2) & = &  \int_0^1 dx \,
                              [ \tilde{g}   (x)
                              + \tilde{f_1} (x) ]
  \nonumber \\
    h_3(m_1,m_2,m_3;k^2) & = &  \int_0^1 dx \, 
                              [ \tilde{g}   (x)
                              + \tilde{f_1} (x) ] \, (1-x)
  \nonumber \\
    h_4(m_1,m_2,m_3;k^2) & = &  \int_0^1 dx \,
                              [ \tilde{g}   (x)
                              + \tilde{f_1} (x)
                              + \tilde{f_2} (x) ]
  \nonumber \\
    h_5(m_1,m_2,m_3;k^2) & = &  \int_0^1 dx \,
                              [ \tilde{g}   (x)
                              + \tilde{f_1} (x)
                              + \tilde{f_2} (x) ] \, (1-x)
  \nonumber \\
    h_6(m_1,m_2,m_3;k^2) & = &  \int_0^1 dx \,
                              [ \tilde{g}   (x)
                              + \tilde{f_1} (x)
                              + \tilde{f_2} (x) ] \, (1-x)^2
  \nonumber \\
    h_7(m_1,m_2,m_3;k^2) & = &  \int_0^1 dx \,
                              [ \tilde{g}   (x)
                              + \tilde{f_1} (x)
                              + \tilde{f_2} (x)
                              + \tilde{f_3} (x) ]
  \nonumber \\
    h_8(m_1,m_2,m_3;k^2) & = &  \int_0^1 dx \,
                              [ \tilde{g}   (x)
                              + \tilde{f_1} (x)
                              + \tilde{f_2} (x)
                              + \tilde{f_3} (x) ] \, (1-x)
  \nonumber \\
    h_9(m_1,m_2,m_3;k^2) & = &  \int_0^1 dx \,
                              [ \tilde{g}   (x)
                              + \tilde{f_1} (x)
                              + \tilde{f_2} (x)
                              + \tilde{f_3} (x) ] \, (1-x)^2
  \nonumber \\
    h_{10}(m_1,m_2,m_3;k^2) & = &  \int_0^1 dx \,
                              [ \tilde{g}   (x)
                              + \tilde{f_1} (x)
                              + \tilde{f_2} (x)
                              + \tilde{f_3} (x) ] \, (1-x)^3
   \; \; .
\label{hfunctions}
\end{eqnarray}
The four building blocks $\tilde{g}(x)$, $\tilde{f_1}(x)$, $\tilde{f_2}(x)$,
and $\tilde{f_3}(x)$  of these one-dimensional integral representations
are explicitly given in section \ref{sec:method}.

\section*{Appendix 5: Anomalous dimensions}

We quote here all anomalous dimensions necessary for the 
calculation. They were presented in \cite{MisiakBobeth,Paolonew} and 
have been checked by us. 

The counterterm contribution which are  proportional to $C_1$ and $C_2$  due to the mixing of the two operators 
${\cal O}_1$ and ${\cal O}_2$ with the operators 
${\cal O}_j \quad (j=1...12)$ are given by the following matrix elements  
\begin{equation}
 \langle s\ell^+\ell^-| \, \sum_{j=1}^{12} \delta Z_{ij}  O_j  \, | b \rangle \, , 
\quad i=1,2 ,
\end{equation}
with the renormalization constants $Z_{ij} = \delta_{ij} + \delta Z_{ij}$ 
where 
\begin{equation}
    \delta Z_{ij} = \frac{\alpha_s}{4 \pi} \left( a_{ij}^{01} +
\frac{1}{\epsilon} a_{ij}^{11}\right) +
        \frac{\alpha_s^2}{(4 \pi)^2}
        \left( a_{ij}^{02} + \frac{1}{\epsilon} a_{ij}^{12} +
\frac{1}{\epsilon^2} a_{ij}^{22}\right) +
        O(\alpha_s^3).
\label{operatorcounterterm}
\end{equation}
There are two evanescent operators ${\cal O}_{11}$ and ${\cal O}_{12}$ 
involved in the mixing 
in addition to the operator basis given in (\ref{heffll}). 
As usual, their choice is not unique, but we follow 
the ones used by  the authors of 
\cite{MisiakBobeth} so as  to be able to use their 
results for the Wilson coefficients: 
\begin{eqnarray}
   {\cal O}_{11} &=& ( {\bar s}_L \gamma_\mu \gamma_\nu \gamma_\sigma T^a c_L )
             ( {\bar c}_L \gamma^\mu \gamma^\nu \gamma^\sigma T^a b_L ) -
16 \,\,  {\cal O}_1\, ,   \nonumber\\
    {\cal O}_{12} &=& ( {\bar s}_L \gamma_\mu \gamma_\nu \gamma_\sigma c_L )
             ( {\bar c}_L \gamma^\mu \gamma^\nu \gamma^\sigma b_L ) - 16
\,\,  {\cal O}_2 \, .
\end{eqnarray}
The anomalous dimensions are given then by:

\begin{table}[htb]
\begin{center}
\begin{tabular}{| c | c | c | c | c | c | c | c | c | c | c | c | c |} \hline 
$a^{11}_{ij}$  & j=1  &  j=2 &j=3&j=4&j=5&j=6&j=7&j=8&j=9&j=10&j=11&j=12\\ \hline
i=1 & -2 & 4/3 & 0 & -1/9 & 0 & 0 & 0 & 0& -16/27 & 0 & 5/12 & 2/9\\ \hline
i=2  & 6 & 0 & 0& 2/3 & 0 & 0 & 0&  0& -4/9 & 0 & 1 & 0 \\ \hline
\end{tabular}
\end{center}
\end{table} 
\begin{table}[htb]
\begin{center}
\begin{tabular}{| c | c | c | c | c | c | c | c | c | c | c | c | c |} \hline 
$a^{22}_{ij}$  & j=1  &  j=2 &j=3&j=4&j=5&j=6&j=7&j=8&j=9&j=10&j=11&j=12\\ \hline
i=1 & * & * & * & * & * & * &  * &  * & 1168/243 & *  &  * &  *\\ \hline
i=2  & * & * & * & * & * & * & *   &  * & 148/81 & *  & * & * \\ \hline
\end{tabular}
\end{center}
\end{table} 
\begin{table}[htb]
\begin{center}
\begin{tabular}{| c | c | c | c | c | c | c | c | c | c | c | c | c |} \hline 
$a^{12}_{ij}$  & j=1  &  j=2 &j=3&j=4&j=5&j=6&j=7&j=8&j=9&j=10&j=11&j=12\\ \hline
i=1 & * & * & * & * & * & * & -58/243 &  * & -64/729 & *  &  * &  *\\ \hline
i=2  & * & * & * & * & * & * & 116/81  &  * & 776/243 & *  & * & * \\ \hline
\end{tabular}
\end{center}
\end{table}

\section*{Appendix 6:  $O(\Lambda^2_{\rm QCD}/m_c^2)$ corrections. }
The explicit expressions of the $O(\Lambda^2_{\rm QCD}/m_c^2)$
non-factorizable corrections  to $R(s)$  and $A_{\rm FB}$ are \cite{Rey}:
\bea
\delta_{1/m^2_c} R(s)&=&\frac{8 \lambda_2}{9 m^2_c} 
\frac{\alpha_{\rm em}^2}{4\pi^2} \frac{V_{cs}^* V_{ts} }{V_{cb}^*V_{tb}}
\frac{(1-s)^2}{f(z)\kappa(z)}\ \Re \Biggl[ 
\frac{1+6s-s^2}{s}  F\left( \frac{s}{4z}\right) C_2 C^{\rm new}_7(s)^*  \no \\
&& \qquad\qquad\qquad\qquad\qquad\qquad\quad 
   + (2+s)  F\left( \frac{s}{4z}\right) C_2  C^{\rm new}_9(s)^* \Biggl]~, \\
\delta_{1/m^2_c} A(s)&=& - \frac{\lambda_2}{3 m^2_c} \frac{\alpha_{\rm em}^2}{4\pi^2}
\frac{V_{cs}^* V_{ts} }{V_{cb}^*V_{tb}}
\frac{(1-s)^2}{f(z)\kappa(z)}\ \Re \Biggl[  ( 1+3s)  F\left( \frac{s}{4z}\right)
C_2 C^{\rm new}_{10}(s)^* \Biggl]~,
\eea
where 
\be
\label{frl1}
F(r)=\frac{3}{2r}\left\{ \begin{array}{ll}
\displaystyle\frac{1}{\sqrt{r(1-r)}}\arctan\sqrt{\frac{r}{1-r}} 
   -1 &  \qquad\qquad 0< r < 1~, \\
 \displaystyle\frac{1}{2\sqrt{r(r-1)}}\left(
\ln\frac{1-\sqrt{1-1/r}}{1+\sqrt{1-1/r}}+i\pi\right)-1 &
\qquad\qquad r > 1~. \end{array} \right. 
\ee

\bigskip

\frenchspacing
\footnotesize
\begin{multicols}{2}

\end{multicols}
\end{document}